\documentclass[aip,jcp,a4paper,reprint,onecolumn]{revtex4-1}
\usepackage{amsmath}
\usepackage{amssymb}
\usepackage[dvips]{graphicx}
\usepackage{color}
\usepackage{tabularx}
\usepackage{gensymb}

\newcommand{\vect}[1]{\textbf{\textit{#1}}}
\newcommand{\dd}[1]{\textrm{d}}

\newcommand{\Atom}{{\textrm{{Atom}}}}

\newcommand{\CG}{{\textrm{CG}}}
\newcommand{\HY}{{\textrm{$\Delta$}}}

\usepackage{algorithm}
\usepackage{algorithmic}

\begin{document}
%--------------------------------------------------------------------
%-----------------------------------------------------------------
\title{Adaptive Resolution Simulation as a Grand Canonical Molecular Dynamics Scheme: Principles, Applications and Perspectives}
\author{Luigi Delle Site}
\email{luigi.dellesite@fu-berlin.de}
\affiliation{Institute for Mathematics, Freie Universit\"at Berlin, Germany}
\author{Animesh Agarwal}
\email{animesh@zedat.fu-berlin.de}
\affiliation{Institute for Mathematics, Freie Universit\"at Berlin, Germany}
\author{Christoph Junghans}
\affiliation{Theoretical Division (T-1), Los Alamos National Laboratory, Los Alamos, NM 87545, USA}
\email{junghans@lanl.org}
\author{Han Wang}
\email{han.wang@fu-berlin.de}
\affiliation{Institute for Mathematics, Freie Universit\"at Berlin, Germany}

\begin{abstract}
We describe the adaptive resolution multiscale method AdResS. The conceptual evolution as well as the improvements of its technical efficiency are described step by step, with an explicit reference to current limitations and open problems.
\end{abstract}

\maketitle

\section{Introduction}
Many problems in the field of condensed matter, intended in the broadest sense, are multiscale in nature.
The concept of ``multiscale'' is here intended as the interplay between
different scales, where such an interplay must have the fundamental role in the properties of a system.
Unfortunately nowadays the term ``multiscale'' is often misused, and, above all in the community of molecular simulation,
almost every problem is forced into this category.
In reality, an exhaustive description of physical phenomena corresponding to a multiscale process requires, in
principle, the simultaneous treatment of all the degrees of freedom (DOFs) of the system. This is obviously a
prohibitive task not only because of the required computational resources but above all because
of the large amount of data produced. In fact, we would have to deal with a huge amount of data that would  mostly contain 
information not essential to the problem analyzed and will inevitably overshadow the
underlying fundamental physics or chemistry of the system.
In general, the desired goal is that of treating in a simulation only
those DOFs which are strictly required by the specific problem  and the choice of the relevant DOFs should be done in such a way that fundamental physical principles are not violated.
This aim is at the basis of the modern research industry dedicated to the development of multiscale models in simulation, however, too often, the basic requirement of physical conceptual consistency is violated.
From the practical point of view, in order to produce methods that allow us to study large scale phenomena, physical approximations are often unavoidable, however one must be really careful regarding this aspect. In fact it should be mandatory to address the following points: (a) why some methods work particularly well, despite conceptual conditions are not met; (b) a systematic determination of the error made because of the physical approximations done.
Most of the multiscale methods in molecular simulation do not proceed along these directives, thus, despite their popularity, they do not gain credibility outside the field of molecular simulation.\\
In this perspective, this chapter will report about the development and application of, AdResS \cite{jcp,prl1,prx}, a multiscale method based on the concept of adaptive molecular resolution in space. We will report its theoretical foundations, its merits, its limitations and its extension as a Grand Canonical Molecular Dynamics set up.
Large space will be given to the discussion of why the method sometimes works and sometimes does not, to the necessity of providing, as much as possible, a mathematical/rigorous basis to the theoretical foundations and to the capability of defining criteria of control of errors caused by physical approximations. Our aim is to provide the reader with the tools for understanding the current level of conceptual consistency and the capability of controlling of approximations/errors; finally, several sections are dedicated to open problems where further development is still required.
\section{Theoretical foundations}
The AdResS approach falls in the general category of {\it concurrent coupling methods} \cite{entropy}, that is, those methods where scales cannot be separated and all the relevant DOFs  must be treated simultaneously. In order to slowly introduce the idea we make here the similarity with the well established Quantum Mechanics/Molecular Mechanics (QM/MM) method \cite{Laio:2002}. QM/MM is based on the idea that a fixed subsystem
is described with a quantum resolution while the remaining part of the system is
treated at classical atomistic level, in this book there are several examples of such an approach. The similarity of QM/MM with AdResS, is that space is divided in different regions in each of which a different molecular resolution is employed according to the requirements of the problem. However, {\it per se}, the QM/MM approach leads to several conceptual problems, the relevant one for the discussion here is that the two regions are not open, thus the free exchange of molecules from one region to the other is forbidden and thus particles density fluctuations are automatically suppressed; these latter are crucial for the correct description of the thermodynamics of bio-systems. Of course one can take into account this aspect and quantify the possible error, however the predictive power of QM/MM is highly restricted when the interplay between thermodynamic fluctuations and electronic structure is a key aspect of a physical process. The idea of AdResS is instead that of allowing for the free exchange of particles, which, passing from one region to the other, change their resolution accordingly and preserve the thermodynamics of the system, that is the thermodynamic properties the system would have if it was treated as a whole at high resolution. The process of changing resolution and preserve the thermodynamic (and structural) properties requires the development of statistical mechanics concepts which do not follow the standard textbook guidelines and thus are often subject to the prejudice of the large portion of conservative researchers in the molecular simulation community.
In the next sections we will describe, step by step, the concept we have developed, starting from an empirical intuitive idea up to the definition of a Grand Canonical-like set up. It must be underlined, that differently from QM/MM, at this stage, we couple only classical (atomistic and coarse-grained) systems, however, idea based on the AdResS scheme have been, in the meanwhile, employed to extend QM/MM to an open system set up. A brief discussion about this aspect can be found in the section dedicated to the extension of AdResS to quantum systems.

\subsection{From AdResS to GC-AdResS}
The starting point of the original AdResS idea is an intuitive empirical principle whose main components are:
\begin{itemize}
\item the simulation box is divided in three regions, one characterized by atomistic resolution, one at coarse-grained (e.g. spherical) resolution and, in between, a small transition region where molecule have a hybrid resolution as explained below.
\item Molecules in the different regions are coupled through a spatial interpolation formula on the forces:
\begin{equation}
{\bf F}_{\alpha \beta}=w(X_\alpha)w(X_\beta){\bf
  F}_{\alpha\beta}^{Atom}+[1-w(X_\alpha)w(X_\beta)]{\bf F}^{CG}_{\alpha\beta} 
\label{eqforce}
\end{equation}
where $\alpha$ and $\beta$ indicate two molecules, ${\bf F}^{Atom}$ is the
force derived from the atomistic force field and  ${\bf F}^{CG}$
from the corresponding coarse-grained potential, $X$ is the $x$ coordinate of the center of mass of
the molecule and $w$ is an interpolating function which smoothly goes from $0$
to $1$ (or vice versa) in the interface region, ($\Delta$), where the lower resolution is 
slowly transformed (according to $w$) in the high resolution (or vice versa),
as illustrated in Fig.\ref{fig1}.\\
\begin{figure}
\center
\includegraphics[width=0.475\textwidth]{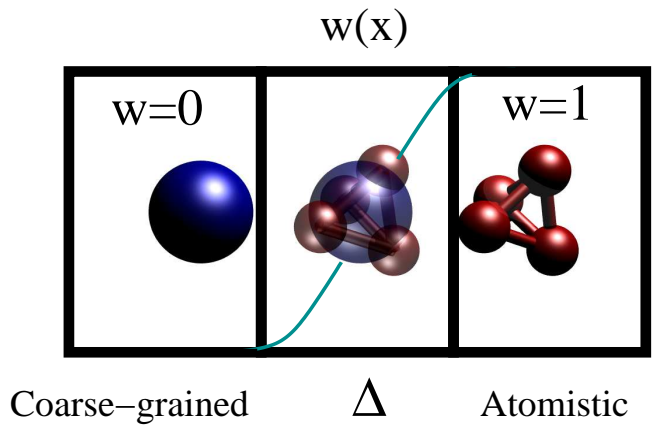}
\caption{Pictorial representation of the adaptive box and
corresponding molecular representation. The region on the left,
is the low resolution region (coarse-grained), the central part is
the transition (hybrid) region $\Delta$, where the switching function
$w(x)$ is defined, and the region on the right, is the high resolution region (atomistic).\label{fig1}}
\end{figure}
According to Eq.\ref{eqforce} two atomistic molecules interact via a force derived only from an atomistic potential, coarse-grained molecules, instead, interacts with all the others via a  force derived only via a coarse-grained potential (coarse-grained molecules do not posses any atomistic degrees of freedom), while for the other cases molecules interact according to their coupled value of $w(X_\alpha)w(X_\beta)$, that is with some space-dependent hybrid combination of atomistic and coarse-grained potential . This implies that a molecule moving from the atomistic to the coarse-grained region, experience a slowly diminishing effect of the force on its atomistic degrees of freedom (rotations and vibrations) and experience an increasing effect (on the center of mass) of the coarse-grained potential. This implies that the molecule experiences the various interactions more and more as a sphere would do. The same process, in reverse order, (acquire degrees of freedom) happens to a coarse-grained molecule going towards the atomistic region.
A detailed explanation of the technical implications of Eq.\ref{eqforce} can be found in \cite{jungsim,jctc} . 
The (physical) empirical, idea behind Eq.\ref{eqforce} is that a smooth interpolation on the forces will allow for a transition from an atomistic to a coarse-grained dynamics of a molecule when it goes from the atomistic to the coarse-grained region (and vice versa). Since the interpolation is smooth and the transition region is relatively small compared to the other regions, the perturbation of the dynamics in the atomistic or coarse-grained region, due to the change of resolution, is negligible and thus the behavior of molecules in the two regions is statistically very close to that of a full atomistic or full coarse-grained simulation (in each corresponding region).  
\item A thermostat is added to assure the overall thermodynamic equilibrium at a given temperature and to take care of the loss of energy due to the perturbation of the transition region.
\end{itemize}
This essential set up, though very simplified, turned out to be extremely accurate when implemented in an MD approach for toy systems \cite{jcp,prefirst,jcppol}; later it was successfully employed to simulate physio-chemical systems \cite{wat1,wat2,jcpcover}. 
The coarse-grained model was derived from full atomistic simulations in such a way that it reproduces the molecule-molecule radial distribution  at the same temperature, density and pressure \cite{ibi,votca}.
However in these studies it was noted that in the transition region there was a small systematic discrepancy between the target particle number density and that obtained in the adaptive simulation.
It is obvious that in this region the accuracy cannot be high but on the other hand this region is of no physical interest for the simulation. However the discrepancy mentioned above, implies that also in the atomistic and coarse-grained region there is a systematic error (though small); we will show in later sections that reproducing the density in the transition region is indeed crucial for the Grand-Canonical-like set up. The natural question arising at this point is why such a discrepancy occurs and whether it is possible to develop a scheme to control and improve the accuracy in the transition region. If such a scheme can be found then as a consequence we would be able to improve the accuracy in the region we are mostly interested, that is the high resolution (atomistic) region.  The question posed above implies a deeper understanding of the thermodynamics and statistical mechanics of a system simulated with adaptive molecular resolution and the essence of the problem can be summarized as it follows:
\begin{itemize}
\item The adaptive approach is based on force interpolation~\eqref{eqforce} and such interpolated force cannot be derived from a potential (for a rigorous proof see \cite{jpa,preluigi}); this aspect will be discussed in a detailed way in one of the next sections.
\item The lack of a global Hamiltonian implies that thermodynamic properties cannot be defined on the basis of standard statistical mechanics. The \emph{laws} of standard thermodynamics should be reconsidered and properly introduced within the framework of the adaptive resolution system.  
\item Given the scenario reported above, it naturally comes the question of how, when using AdResS, one can assure {\it a priori} that thermodynamic and structural properties are properly described.
\item However, {\it a posteriori} checks of thermodynamics and structural properties of simulated systems, show that the accuracy is actually very high. The only sizeable discrepancy is the previously reported enhanced value of the particle density in the transition region.
\end{itemize}  
The approach employed to solve the question above consists of going beyond the traditional path of defining (at any cost) a Hamiltonian. This, in fact, would imply that we are likely to perform the calculation of relevant thermodynamic properties in a ``forced/artificial'' Canonical ensemble where artifacts may influence results in a sizeable manner. The question to pose is not about the strict definition of the statistical ensemble, rather the definition of the thermodynamic quantities of interest and how to properly calculate them in a molecular simulation characterized by the process of adaptive molecular representation. This is actually the key point, the adaptive resolution does not have any physical meaning, a system of $N$ water molecules must have the overall properties of water independently from the molecular representation. It will have a more detailed local description in the atomistic region, but its overall density, temperature and pressure must be the same everywhere (atomistic, coarse-grained and even transition region) and should be the same as those of a reference full atomistic system of $N$ molecules. With this principle in mind, two key steps were done; the first regards a rigorous definition of temperature in the transition region. 
\subsubsection{Temperature in $\Delta$}
Temperature is well defined in the atomistic and coarse-grained region via the equipartition theorem, as the average kinetic energy divided by the total number of DOFs, $T_{Atom}=\frac{2\left <K_{Atom}\right>}{n_{Atom}}$;
$T_{CG}=\frac{2\left <K_{CG}\right>}{n_{CG}}$. 
However, such a definition does not apply in the transition region where one would have: $T_{\Delta}=\frac{2\left <K_{\Delta}\right>}{n_{\Delta}}$; here DOFs are partially removed/reintroduced as a function of the position thus one cannot properly define $\left<K_{\Delta}\right>$, while $n_{\Delta}=n_{Atom}w(x) + n_{CG}(1 - w(x))$, with $n$ the number of DOFs, molecules would have if they had full atomistic resolution. At this point, one must notice that as a matter of fact, the calculation of the average kinetic energy in an ideal canonical ensemble characterized by an adaptive molecular resolution, would imply that in the transition region each removed/reintroduced DOF does not fully contribute to the average kinetic energy. In fact in the integral of $\left<K\right>$ the dimensionality of a transforming DOF goes between zero (a coarse-grained molecule does not have this DOF) and one (an atomistic molecule fully possesses such a DOF). 
This situation can be described in an exact way via the mathematical formalism of integration in fractional dimensions and implies only to consider an extra multiplicative term in the integral of $\left<K\right>$ \cite{jpa,prefrac}. The overall result shows that the temperature in each point, $x$ in the $\Delta$ region can be calculated as: $T_{\Delta}=\frac{2\left <K_{\Delta}\right>}{n_{\Delta}(x)}$; moreover it turned out that $\left<K_{\Delta}\right>$ is proportional to $n_{\Delta}(x)$ and thus the ratio, $\frac{\left<K_{\Delta}\right>}{n_{\Delta}(x)}$, leads to a constant value which is nothing else than the target temperature of the reference full atomistic system. Computational tests showed that indeed, with the definition of temperature in $\Delta$, as derived above, the system was automatically kept at the expected thermal equilibrium during the simulation. This was the first essential step in the direction of having a deeper understanding and justification of the basic algorithm. Once the procedure for the calculation of the temperature was proven satisfactory, then we moved to next problem, that is address the question of why in $\Delta $ a small, but systematic, discrepancy for the particle density is observed in all simulations. 

\subsubsection{Balancing the chemical potential}
Let us consider the typical AdResS set up and ask the question of what would preserve thermodynamic equilibrium, and thus uniform molecular density, if one interfaces three open systems with different molecular representations. If each molecular resolution was a different molecular species and, in first instance, one starts with the same density all over the space, then the different species would remain in equilibrium if the chemical potential is the same for all species. For an adaptive set up, molecules have only a different representation, they do not belong to different species and are free to move from one region to the other, however, for similarity to the case above, if the density must remain constant, then the chemical potential characterizing each molecular resolution (at the given temperature, density and pressure) should be the same as that of the reference full atomistic case. Let us suppose it is possible to calculate the chemical potential corresponding to each resolution, this would lead, over the simulation box, to a space dependent chemical potential $\mu(x)$, where $\mu(x)=\mu_{Atom}$ in the atomistic region, $\mu(x)=\mu_{CG}$ in the coarse-grained region and $\mu(x)=\mu_{\Delta}(x)=\mu(w(x))$ in the transition region. Next, taking the atomistic resolution as reference, it would be sufficient to balance the chemical potential in each region, that is, one can impose that: $\mu^{eff}=\mu_{Atom}$, ($\mu^{eff}$ is the effective chemical potential that characterizes the entire system) independently from the resolution and thus of the position, then we would have equilibrium and uniform density all over the simulation box. This target can be obviously accomplished if one imposes a force proportional to the gradient of the difference between the reference chemical potential and the space dependent chemical potential: ${\bf F}(x_{\Delta})\approx\nabla[\mu_{Atom}-\mu(x)]$. Such a force should act on the center of mass of the molecules in $\Delta$, and automatically vanishes in both atomistic and coarse-grained regions.\\
While the calculation of $\mu_{Atom}$ and $\mu_{CG}$ can be done in simulation via thermodynamic integration or Widom, {\it insertion particle} approach \cite{tint,ipm} by considering a full atomistic system and a full coarse-grained system, the question arises on how to calculate $\mu(w(x))$ in $\Delta$, in fact such techniques require the definition of a Hamiltonian. To this aim, a scheme was developed where $\Delta$ was ideally divided in several small regions each characterized by a constant value of $w$. Then each constant value of $w$ was considered separately, that is, an auxiliary simulation was performed where all the molecules in the simulation box interact with an AdResS force with $w$ fixed to a given value (not anymore space dependent). In such a set up it is possible to define a Hamiltonian (corresponding to the linear interpolation via the constant $w$ of atomistic and coarse-grained potential) and thus calculate the chemical potential, $\mu(\hat{w})$ corresponding to the specific value of $w=\hat{w}$.
Since each constant value of $w$ characterizes a small region of $\Delta$, $\mu(w)$, calculated in the way described before, is an approximate, though numerically accurate, way to determine $\mu(x)$ in $\Delta$. Next, one can take a force proportional to the gradient, ${\bf F}_{th}\approx\nabla[\mu_{Atom}-\mu(x)]$, called then ``{\it thermodynamic force}'', apply it to the center of mass of the molecules and, as a consequence, a uniform molecular density should be obtained while preserving all other thermodynamic and structural properties (temperature, pressure, radial distribution functions, solvation structures, etc. etc.). This idea was proved to be successful not only for standard liquids, but also for liquid mixtures \cite{jcpmu}.
The key point is that there was no need to explicitly define a global (most probably artificial) Hamiltonian but yet the basic properties were accurately reproduced employing conclusions based on thermodynamics only. Despite this study provided a deeper understanding of how the process of adaptivity can be interpreted on the basis of standard thermodynamics, from the practical point of view, the calculation of the thermodynamic force required an excessive number of extra simulations. As a consequence this approach, conceptually satisfactory, was not optimal from the practical point of view, however it would stimulate further conceptual developments, as reported below.
\subsubsection{Thermodynamics and Statistical mechanics of a generic open system}
The approach employed above, requires that the atomistic model and the coarse-grained model, at given density and temperature, are characterized by the same pressure. A coarse-grained potential, derived only by the inversion of the molecule-molecule radial distribution function, would preserve the compressibility, but not the pressure. For this reason, usually a pressure correction term is added to the coarse-grained model at the price of loosing the correct compressibility \cite{hanpaper}. However, starting from the basic idea of the previous section, that is obtaining a uniform density also in the transition region, one may think of a generalized principle that imposes a uniform density profile across the simulation box without correcting the pressure of the coarse-grained model.
Said in these terms, this is the intuitive/practical idea, however one can formalize it on the basis of principle of thermodynamic and statistical mechanics.
In fact, the adaptive set up consists of interfaced open systems; from the thermodynamic and statistical mechanics point of view, the key quantity of interest is the grand potential, $\Omega=PV$. The volume of each subsystem is a fixed quantity, as a consequence the equilibrium between the atomistic and the coarse-grained region is given by the equality of pressure. Thus in the interface region, $\Delta$, conditions should be defined such that $p_{Atom}=p_{CG}$ is assured;
this was done via a reformulation of the thermodynamic force in terms of balancing the pressure \cite{prl1}:
\begin{equation}
  p_{Atom}+\rho_{0}\int_{\Delta}{\bf F}_{th}({\bf r})d{\bf r}=p_{CG},
\end{equation}
where $p_{Atom}$  is the target pressure of the atomistic system (region), $p_{CG}$ is the pressure of the coarse-grained model, $\rho_{0}$ is the target molecular density of the atomistic system (region). In the MD set up, ${\bf F}_{th}$ is calculated via an iterative procedure as a function of the molecular density in $\Delta$. The iterative formulas at the $i$-th step is
\begin{equation}
  {\bf F}^{i+1}_{th}({\bf r})={\bf F}^{i}_{th}({\bf r})
  - \frac{1}{\rho_{0}\kappa^2_{T}}\nabla \rho^{i}({\bf r}),
\label{kappa}
\end{equation}
where $\kappa_{T}$ the isothermal compressibility. The force is considered converged when the target density $\rho_{0}$ in $\Delta$ is obtained. In principle, this derivation implies a Grand-Canonical-like set up, in which each region acts as a reservoir for the other and the transition region as a filter that assures the correct entrance/exit of molecules from one region to the other. The approach as it is, cannot be defined Grand-Canonical because of the finiteness of the reservoir(s) and because one cannot calculate directly the chemical potential, $\mu$; this can be only calculated in a separate way at high computational costs, as underlined in the previous section. Nevertheless this approach turned out to be satisfactory from both practical and conceptual point of view \cite{prl1,debashish1,debashish2}, and represented the basis for a further development toward a proper Grand-Canonical-like set up \cite{jctc}, this is reported in the next section.
\subsubsection{Grand-Canonical-like formulation and its mathematical necessary conditions}
\begin{figure}
\center
\includegraphics[width=0.6\textwidth]{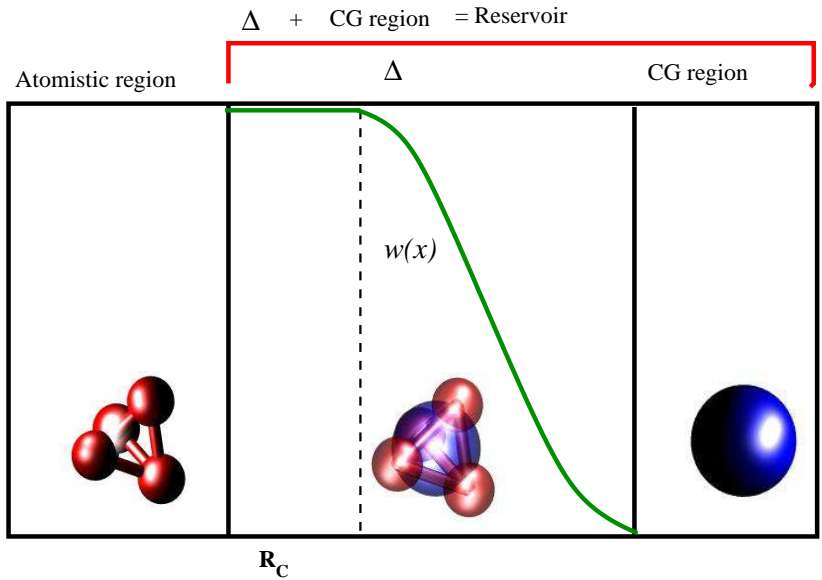}
\caption{Pictorial representation of the new adaptive box. $w(x)$ is extended into the atomistic region of an amount equal to $R_{c}$, the cut off distance of the atomistic interaction.  This enables us to write an exact Hamiltonian for the atomistic region and identify the rest of the system as a generic reservoir.}\label{fig:adapt-extended}
\end{figure}

\begin{figure}
\center
\includegraphics[width=1.05\textwidth]{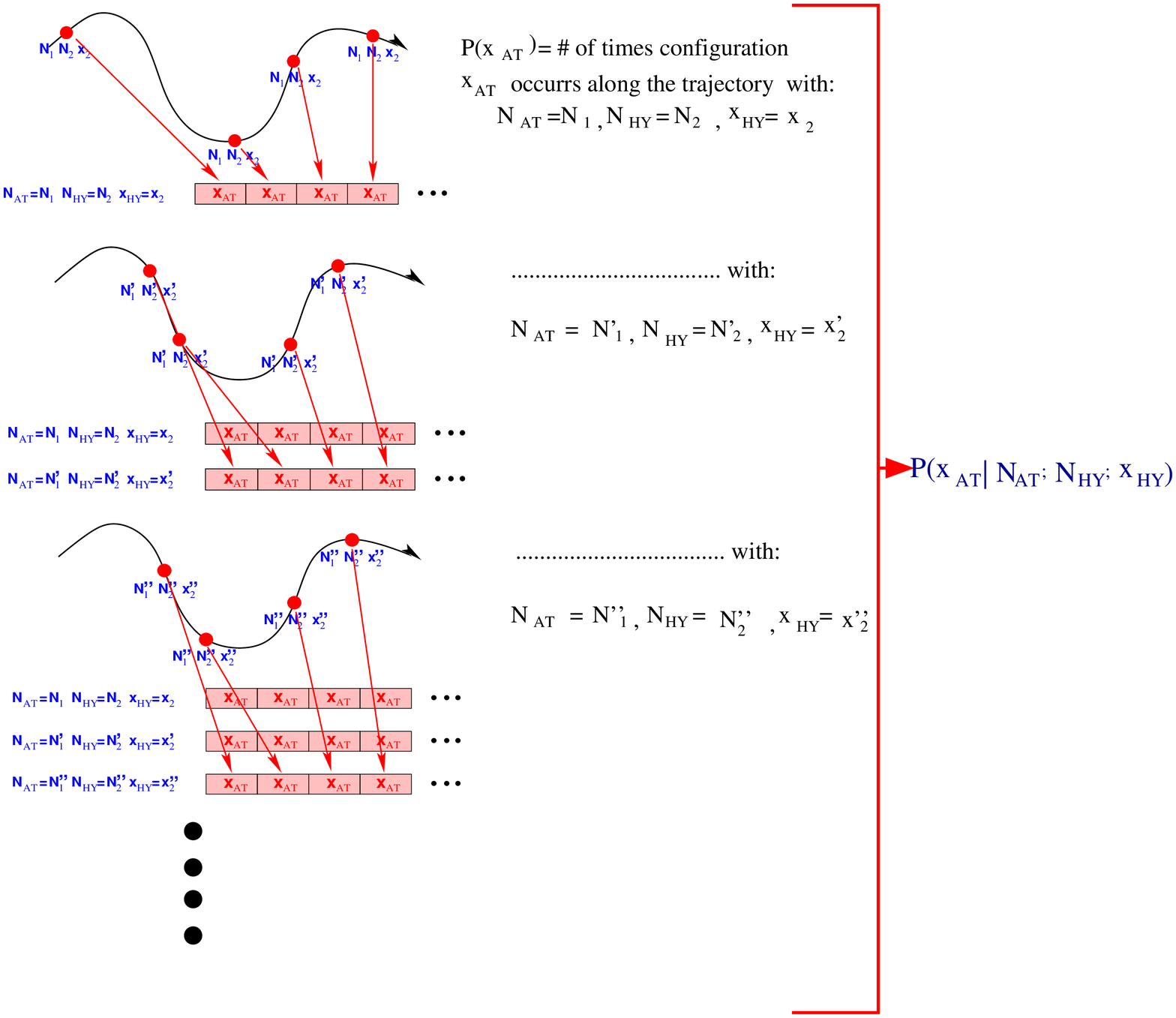}
\caption{Pictorial representation of the ``conceptual'' process of how to derive $P(x_{Atom}|N_{Atom};N_{\Delta}; x_{\Delta})$ from a generic trajectory of AdResS.
First, one ideally fixes $N_{Atom}$ (here indicated by $N_{AT}$), $N_{\Delta}$ (here indicated by $N_{HY}$) and $x_{\Delta}$ (here indicated by $x_{HY}$), and counts the number of times a given configuration $x_{atom}$ (here indicated by $x_{AT}$) occurs at under this condition. This is (ideally) repeated for all $x_{Atom}$.
Let us suppose that the parameters chosen at this first step are: $N_{AT}=N_{1};N_{HY}=N_{2}; x_{HY}=x_{2}$, then the process of sorting out of configurations under this condition leads to $P(x_{Atom}|N_{1};N_{2}; x_{2})$. At next step the parameters chosen are: $N_{AT}=N^{'}_{1};N_{HY}=N^{'}_{2}; x_{HY}=x^{'}_{2}$ and the sorting out process for all $x_{AT}$ is repeated and one obtains: $P(x_{Atom}|N^{'}_{1};N^{'}_{2}; x^{'}_{2})$. Next, a new sets of parameters  are defined: $N_{AT}=N^{''}_{1};N_{HY}=N^{''}_{2}; x_{HY}=x^{''}_{2}$, and one obtains: $P(x_{Atom}|N^{''}_{1};N^{''}_{2}; x^{''}_{2})$. When the process is iterated at infinity, the collection of $P(x_{Atom}|N^{i}_{1};N^{i}_{2}; x^{i}_{2})$ (with $i$ generic step), corresponds to: $P(x_{Atom}|N_{Atom};N_{\Delta}; x_{\Delta})$. If the AdResS trajectory is, in principle, infinitely long,one can imagine that for each choice of the parameters $N_{AT};N_{HY}; x_{HY}$ the statistical sorting out is equivalent to run a simulation with a Hamiltonian where $N_{AT};N_{HY}; x_{HY}$ are fixed, and thus such probability has a Boltzmann form and as a consequence this holds also for $P(x_{Atom}|N_{Atom};N_{\Delta}; x_{\Delta})$.\label{fig-sequence}}
\end{figure}

The approach reported above leads, in a natural way, to the question of how much AdResS can be formulated in terms of Grand-Canonical ensemble and thus obtain a statistical mechanics framework that justifies, from the conceptual point of view, its computational robustness.
For this purpose we must address the key question whether or not the probability
distribution of the atomistic phase space variables $\vect x_\Atom$ and
the number of molecules in the atomistic region $N_\Atom$ are subject to
the Grand-Canonical distribution.
This implies the possibility of writing and numerically verifying the following formula:
\begin{align}\label{eqn:gc-dist}
  p(\vect x_\Atom, N_\Atom) = \frac{1}{\mathcal Z_\Atom}
  e^{\beta\mu_{\Atom} N_\Atom - \beta \mathcal H_{N_\Atom}^{\Atom}(\vect x_\Atom)} 
\end{align}
Here $\mathcal Z_\Atom$ is the grand-canonical partition function and
the $\mathcal H^{\Atom}$ is the atomistic Hamiltonian. In order to properly set the question, we have to consider the case of
thermodynamic limit, where the atomistic region is large enough in order to have a statistical meaning, the transition region works as a relatively thin filter and the coarse-grained region acts as a large reservoir; this assumes
 $N_\CG \gg 1$,
$N_\Atom \gg 1$, $N_\CG
\gg N_\Atom \gg N_\HY$.
However, in practice, surprisingly, the method is characterized by high accuracy even in the ``worst case scenario'', in which the atomistic and CG
regions can be even smaller than the hybrid region~\cite{prx}.  In
this limiting process, as anticipated above, the size of the hybrid region is kept constant,
therefore, it can be viewed as a thin filter that allows the change of resolution when a molecule freely goes from a high resolution
region to a low resolution region and vice versa. In the initial condition set up of the adaptive simulation, different
resolutions (regions) are at different thermodynamic state (different pressure), therefore, to assure
thermodynamic equilibrium of the whole system (to be the same of the target full atomistic simulation), we can require that the
hybrid region performs some works (denoted by $\omega_0$ in this chapter)
when a molecule goes through the hybrid region.
In this section we discuss how such a work should be defined and what is its physical meaning in order to reproduce the target probability density in the atomistic region . Regarding the ``accuracy'' of AdResS, we 
compare the AdResS distribution with that of a full-atomistic simulation. It is clear that in the
thermodynamic limit, a subsystem of a full atomistic system, that is of
the same size as the atomistic region in AdResS, is a natural Grand Canonical
ensemble. Therefore the full-atomistic
reference system serves a good measure of the accuracy of AdResS as a Grand-Canonical MD set up.\\
A direct approach to an analysis of the distribution function~\eqref{eqn:gc-dist}
in an adaptive set up is rather complex, thus we will circumvent the problem through different steps the first of which consists in splitting the distribution as:
\begin{equation}
  p(\vect x_\Atom, N_\Atom) = p(\vect x_\Atom \vert N_\Atom) \,{p (N_\Atom)}
\label{eqn:split}
\end{equation}
As initial step we consider the first term of the R.H.S.~of Eq.~\eqref{eqn:split}:
\begin{equation}
  p(\vect x_\Atom \vert N_\Atom) =
  \sum_{N_\HY} \int \textrm{d}\vect x_\HY\,
  p(\vect x_\Atom \vert N_\Atom; \vect x_\HY, N_\HY)\cdot
  p(\vect x_\HY, N_\HY\vert N_\Atom)
\label{pdelta}
\end{equation}
The conditional distribution is a Boltzmann distribution
\begin{equation}
  p(\vect x_\Atom \vert N_\Atom; \vect x_\HY, N_\HY) \propto
  \exp\{-\beta \mathcal H^\Atom_{N_\Atom}(\vect x_\Atom; \vect x_\HY, N_\HY)\}
\label{eqn:at-boltzmann}
\end{equation}
where
\begin{equation}
  \mathcal H^\Atom_{N_\Atom}(\vect x_\Atom; \vect x_\HY, N_\HY) 
  ={{ \sum_{i=1}^{N_\Atom}\frac12 m_i\vect v_i^2 +
      \sum_{i<j}^{N_\Atom}U^\Atom(\vect r_i - \vect r_j) }}
  + {\sum_{i=1}^{N_\Atom}\sum_{j=N_\Atom+1}^{N_\Atom + N_\HY}U^\Atom(\vect r_i - \vect r_j)}.
\label{eqn:at-h}
\end{equation}
In order to properly define an exact $ \mathcal H^\Atom_{N_\Atom}$ without any artifact, we propose a technical modification of the transition region and of the corresponding definition of the switching function $w(x)$.
In \cite{jctc}, the switching function was redefined in such a way that it allows the atomistic interactions to be separated from hybrid and coarse-grained interactions. This was done by extending the transition region into the original atomistic region for an amount corresponding to the cutoff of the molecule-molecule interaction (see also Fig.~\ref{fig:adapt-extended}). As a consequence all the molecules of the (new) atomistic region interact with the rest of the system only in an atomistic way. The price to pay when extending the transition region is balanced by a significative conceptual advancement: now one can write an exact Hamiltonian for the atomistic region (see Eq.\ref{eqn:at-h}), without any unphysical quantity involved, and consider the rest of the system as a bath/reservoir of particles.
Eq.~\eqref{eqn:at-boltzmann}, is a valid approximation because the
DOFs in the hybrid region, $\vect x_\HY, N_\HY$, can be viewed in terms of frozen configuration frames
which, although cannot be treated as deterministic dynamical sequences within AdResS, can be treated in terms of statistical distribution of configurations (see the pictorial explanation of  Fig.\ref{fig-sequence}). 
Therefore, according to Eq.~\eqref{pdelta}, the necessary
condition for having high accuracy for the distribution $p(\vect x_\Atom \vert N_\Atom)$ implies high
  accuracy of the distribution $p(\vect x_\HY, N_\HY\vert N_\Atom)$.
Ideally, it must be the same distribution of the corresponding
subregion of the full-atomistic reference system. In the standard conditions,
this cannot be true, however, one can search for (at least) necessary conditions
which assure a satisfactory accuracy of $p(\vect x_\HY, N_\HY\vert N_\Atom)$.
The condition, $\rho_{\HY} =
\rho_\Atom$, assures a first order accuracy, while the second order is assured by the condition $g_{\HY}(r) = g_\Atom(r)$ .
In a similar way, higher orders of accuracy can be achieved by imposing that 3-body correlation, 4-body correlation functions and so on, are reproduced in $\Delta$, however, in MD, the second order is usually already satisfactory. 
The condition $\rho_{\HY} =
\rho_\Atom$ is fulfilled by the thermodynamic force introduced in the
previous section, while the condition $g_{\HY}(r) = g_\Atom(r)$ can be
fulfilled by a force which corrects the RDF in $\Delta$ to the target one.~\cite{jctc}.
The new expression of the AdResS force acting on a generic molecule $\alpha$ becomes:
\begin{equation}
{\bf F}_{\alpha}=\sum_{\beta\neq\alpha} \left[w_{\alpha}w_{\beta}{\bf F}^{Atom}_{\alpha\beta}+(1-w_{\alpha}w_{\beta}){\bf F}^{CG}_{\alpha\beta}+w_{\alpha}w_{\beta}(1-w_{\alpha}w_{\beta}){\bf F}^{rdf}_{\alpha\beta}\right]+{\bf F}^{th}_{\alpha}
\label{newadress}
\end{equation}
where, $w_{\alpha}w_{\beta}(1-w_{\alpha}w_{\beta}){\bf F}^{rdf}_{\alpha\beta}$, is the corrective force for the radial distribution function,  multiplied by $w_{\alpha}w_{\beta}(1-w_{\alpha}w_{\beta})$ so that it acts only in $\Delta$.
The force ${\bf F}^{rdf}_{\alpha\beta}$ is derived via an Iterative Boltzmann Inversion (IBI) \cite{ibi,votca}, and in this case consists of:
\begin{equation}
U^{rdf}_{i+1}=U^{rdf}_{i}+k_{B}T\ln\left[\frac{g(r)_{i}}{g(r)_{Atom}}\right]
\label{ugr}
\end{equation}
with $U^{rdf}(r)$ being the pairwise correction potential applied to a molecule in $\Delta$, $g(r)_{Atom}$ is the target radial distribution function (full atomistic simulation) and $g(r)_{i}$ is the {\it average} radial distribution function at the $i-th$ step of iteration in $\Delta$. With expression {\it average} here is meant that the $g(r)$ usually depends on the relative position of two molecules, $r={\bf r}_{\alpha\beta}$, however the correction on the $g(r)$ in $\Delta$ in an adaptive scheme, is position dependent, that is some points in $\Delta$ need more/less correction that others. For this reason, in first approximation, we assume that the correction does not have such a dependence in $\Delta$ and we take the average over $\Delta$ so that the correction function is a function of ${\bf r}_{\alpha\beta}$ but uniform for all molecules in $\Delta$.
Next, we construct the prefactor, $w_{\alpha}w_{\beta}(1-w_{\alpha}w_{\beta})$, such that the spatial dependency of the correction in $\Delta$ is reproduced as much as possible, giving to the force the proper weight via $w(x)$. For example, in the region of $\Delta$ closer to the atomistic region, it is needed less correction than in the central region of $\Delta$.
However, since the application of $w_{\alpha}w_{\beta}(1-w_{\alpha}w_{\beta}){\bf F}^{rdf}_{\alpha\beta}$, may perturb the molecular number density in $\Delta$, we perform another iteration for the thermodynamic force, and in turn iteratively correct ${\bf F}^{rdf}_{\alpha\beta}$ until the density and the radial distribution function in $\Delta$ converge both to the target ones.
It is possible to derive also necessary conditions for the accuracy of the second
term of the R.H.S.~of Eq.~\eqref{eqn:split}, i.e.~$p (N_\Atom)$.  The
necessary condition in order to have a first order accuracy is: $\omega_0 =
\mu_{\CG} - \mu_{\Atom}$, that is the work of in $\Delta$ should equalize the difference in chemical potential between the atomistic and the CG resolution.
Instead,  the necessary condition for the accuracy at second
order is: $\kappa_\Atom = \kappa_{\CG}$, that is the equality of the compressibility in the two resolutions. Here the first order
accuracy means that the deviation of $p (N_\Atom)$ from the
full-atomistic reference is controlled by $\textrm{Const}\cdot N_\Atom /
N_\CG$, while the second order accuracy means the deviation is
controlled by $\textrm{Const}\cdot (N_\Atom / N_\CG)^2$. Under the
thermodynamic limit, the deviation vanishes and the
second order decays faster than the first order.  It can be proved that
when the thermodynamic force is imposed to the system, the first order
necessary condition $ \omega_0 = \mu_{\CG} - \mu_{\Atom}$ is satisfied ~\cite{prx} (see also next section for an extended discussion).
The second order accuracy is achieved only when the coarse-grained
model reproduces the atomistic compressibility.

\subsection{Essential quantity in the Grand Canonical ensemble: Automatic calculation of $\mu$} \label{musec}
An essential thermodynamic quantity in the Grand Canonical ensemble is the chemical potential $\mu$. If the framework reported above is properly defined in a Grand Canonical ensemble, then we should be able to determine $\mu$ directly.
In this section we outline the procedure by which $\mu$ can be calculated within AdResS.\\
In Ref.\cite{prx} it was shown that: 
\begin{equation}
\mu_{CG}=\mu_{Atom}+\omega_{th}+\omega_{Q}
\label{mu}
\end{equation}
 $\mu_{CG}$ is the chemical potential of the coarse-grained system, $\mu_{Atom}$  is the chemical potential of the atomistic system and $\omega_{th}=\int_{\Delta}{\bf F}_{th}({\bf r})d{\bf r}$ is work performed by the thermodynamic force in $\Delta$, finally, $\omega_{Q}$ is the heat given by the thermostat by which the inserted/removed degrees of freedom in $\Delta$ are slowly reintroduced/removed. $\omega_{Q}$ consists of two parts, one, $\omega_{extra}$, which takes care of the dissipation of energy caused by the change of interactions in $\Delta$; the other, $\omega_{DOF}$, is related to the equilibration of the reinserted/removed degrees of freedom (e.g. rotational and vibrational). For the equipartition theorem, $\omega_{DOF}=\frac{1}{2}k_{B}T$ per degree of freedom. $\omega_{th}$ can be calculated by integrating the thermodynamic force over $\Delta$. Instead, the calculation of $\omega_{Q}$ is not direct and in Ref.\cite{prx} we have proposed a procedure. The procedure consists of introducing an auxiliary Hamiltonian where the coarse-grained and atomistic potential are now interpolated, instead of forces as in standard AdResS. Next, we impose that the Hamiltonian system has the same thermodynamic equilibrium of the original system considered in a simulation of standard AdResS; this is done by introducing a thermodynamic force which, at the target temperature, assures that the density of particles across the system is the same as in the original system. In this way, in the Hamiltonian approach we obtain the same equilibrium of the original adaptive system  and in addition a thermostat is not needed; this implies that the difference between the work of the original thermodynamic force and the work of the thermodynamic force calculated in the Hamiltonian approach gives $\omega_{Q}$. We have also proven numerically, for liquid water, that $\omega_{Q}=\int_{\Delta}\langle w\nabla_{x} w(x)( V_{AT}-V_{CG})\rangle_{\bf r} d{\bf r}$, with $V_{AT}$ and $V_{CG}$ being the atomistic and coarse-grained potential. The result above means that $\omega_{Q}$ can be calculated in a straightforward way during the  standard AdResS simulation. According to \eqref{mu}, if one knows $\mu_{CG}$, then AdResS automatically gives $\mu_{Atom}$. In MD usually the quantity of interest is not the total chemical potential, but the excess chemical potential $\mu^{ex}_{AT}$, that is, the expression of \eqref{mu} where the kinetic (ideal gas) part is removed. Regarding the kinetic part, the contribution coming from the center of mass for the coarse-grained and for the atomistic molecules is exactly the same and it is automatically removed in \eqref{mu}.
The kinetic part of $\mu_{Atom}$, due to, e.g., the rotational and vibrational degrees of freedom, $\omega_{DOF}$, can be calculated by hand removing $\frac{1}{2}k_{B}T$ per degree of freedom. However, this calculation is actually not required in the current code, since the technical set up of AdResS considers the removed degrees of freedom as phantom variables but nevertheless equilibrate them  in terms of thermal energy\cite{jungsim}. As a consequence the heat given by the thermostat for this part is the same in the atomistic and coarse-graining molecules and it is automatically removed in the difference. Finally, the calculation of $\mu^{ex}_{CG}$ is done with standard methods as Widom insertion particle \cite{ipm} which, for simple coarse-grained models (e.g. spherical molecules) like those of the coarse-grained system, requires a small computational cost.
The final expression writes:
\begin{equation}
\mu^{ex}_{Atom}=\mu^{ex}_{CG}-\int_{\Delta}{F}_{th}({\bf r})d{\bf r}-\int_{\Delta}\langle w\nabla_{x} w(x) (V_{AT}-V_{CG})\rangle_{\bf r} d{\bf r}
\label{finalmu}
\end{equation}
This procedure has been successfully applied to various liquids and mixtures and shows the capability of AdResS to be a general tool for the calculation of thermodynamic properties of a system in a Grand Canonical fashion; for this reason we have renamed this formulation of AdResS as Grand Canonical AdResS (GC-AdResS), however in this paper we will use the name AdResS for any formulation of the method.

\subsection{What about a Hamiltonian approach?}
The strongest criticism to AdResS is not about the principles on which is based, or its computational robustness and performance, but about the fact that the force of Eq.\ref{eqforce} is not conservative, and thus the system does not have a global Hamiltonian. Other authors claim that the force of Eq.\ref{eqforce} can be actually derived from a Hamiltonian where instead of the forces, the atomistic and coarse-grained potential are interpolated in the same technical fashion of AdResS \cite{ensing1,ensing2,ensing3}. However it has been shown both analytically \cite{preluigi} and numerically \cite{prlcomm} that within this scheme, there is no possibility of deriving Eq.\ref{eqforce} from a potential as that suggested in \cite{ensing1,ensing2,ensing3}. The lack of a global Hamiltonian leads to a deep cultural problem: the community of molecular simulation is used to treat systems (in equilibrium) with a well defined Hamiltonian and thus, unfortunately, is rather skeptical regarding procedures where the Hamiltonian is not used, despite they are based (as for AdResS) on well defined first principles of thermodynamics and statistical mechanics. In fact, after the appearance of AdResS in literature, a rush of attempts have been made for defining and adaptive Hamiltonian approach. The one of Ensing {\it et al.} cited above, was essentially equivalent to AdResS but with an inexact Hamiltonian interpretation. A second attempt by the same group is based instead on a Lagrangian approach where the number of particles at high resolution is an explicit part of the equation associated to a Lagrangian multiplier \cite{ensingagain}. Essentially this multiplier is equivalent to a chemical potential which regulates how many particles can be at high resolution, that is, in practice, something equivalent to the thermodynamic force of AdResS.
It follows the question whether GC-AdResS can be written in terms of such a Lagrangian. At this stage we cannot give a final answer, however one must point out the following: the Lagrangian equation of \cite{ensingagain} refers to the description of two open systems (atomistic and coarse-grained) and thus if the term related to the varying number of particles is properly defined, it should imply the existence of a solution for the Liouville equation with varying number of particles (in the atomistic region). Unfortunately, at the current state of the art, the question of the formulation of Liouville equation with varying number of particles is an open problem and actually an exact formulation has not been found yet \cite{arxiv}. It follows that the Lagrangian of 
\cite{ensingagain}, if not associated to a Liouville equation, may be artificial and would give correct results because of an {\it a priori} imposition of what one wants, rather than because of the natural result of a first-principles Lagrangian. Another attempt, to use the Lagrangian approach was developed by Heyden and Truhlar; the idea is based on dividing the space in three regions, as in AdResS, and the region at high resolution is usually centered around a fixed active site; a total Lagrangian is then written as a linear combination of Lagrangians, each corresponding to a possible combination of molecular resolutions and weighted according to the distance from the active center. This means that given a molecule A in the transition region at a distance $r$ from the active center, interacting, for example, with molecules B, C, D, (also in the transition region) the corresponding Lagrangian is a linear combination  of those Lagrangians with all possible combination of resolutions (all molecules have coarse-grained resolution, all have atomistic resolution, A is atomistic and the others are coarse-grained, A is coarse-grained and the others atomistic, A is atomistic, B is atomistic and C and D coarse-grained etc etc). Each of the terms of this  combination is weighted by a function of the distance from the active site, the closer to the atomistic region, the higher the weight to the Lagrangians with more atomistic resolution and vice versa for the coarse-grained. Formally, the result is that one has a Hamiltonian approach and in average certainly takes into account all possible scenarios of transitions from atomistic to coarse-grained and vice versa. However, while formally elegant and robust, it is evident that practically is not feasible, in fact even for a small system of, let us suppose 50 molecules in the transition region, the amount of combinations simply explodes and cannot be calculated. 
In general, if a Hamiltonian approach exists, it is mandatory that satisfies the following necessary requirements:
\begin{itemize}
\item It should lead to the atomistic Hamiltonian in the atomistic region and the coarse-grained Hamiltonian in the coarse grained region
\item It should automatically produce, at least in the atomistic region, a spatial probability distribution as that of the target full atomistic simulation.
\end{itemize}
Recently, Potestio {\it et al.} suggested a global Hamiltonian approach of AdResS \cite{davide}. As for the method of Refs.\cite{ensing1,ensing2,ensing3}, this approach is based on the interpolation in space of the atomistic and coarse-grained potential. The procedure provides an elegant thermodynamic interpretation in terms of Kirkwood's (adiabatic) integration and suggests how to equilibrate the system.
The problem is that similarly to the approach of \cite{ensing1,ensing2,ensing3}, this method by construction, cannot satisfy the first requirement of the necessary conditions above, according to \cite{preluigi}. In fact, the extra force generated by the gradient of the weighting function, $w(x)$ remains the key problem. Such a force induces an unphysical flux of particles from one region to another, the balance of this flux is taken care by an additional field which must be added to the original Hamiltonian, such a field, in \cite{davide}, is derived by an elegant thermodynamic procedure. However, in~\cite{preluigi}  it is shown that such a field (if properly defined) must be a solution of a first order partial differential equation. In order to describe a proper adaptive system, it is required that the field satisfies two boundary conditions; that is, it must be zero in the atomistic and in the coarse-grained region. However, the equation is of the first order and only one boundary condition can be used to fix the solution. For example, let us assume that we fix the boundary condition in the atomistic region; this implies, inevitably, that in the coarse-grained region the original potential is changed by an artificial, unphysical additional term. The consequence is that if we, ideally, reduce the atomistic region to zero, the original coarse-grained potential is not obtained (as instead should be). The additional term is a constant in space, but depends on the data of the specific system. In most of the cases, from the practical point of view, this is a non-relevant aspect, however, from the conceptual point of view, this makes the Hamiltonian, artificial. There is anyway a more relevant aspect; according to \cite{prx}, as it stands now, the method of \cite{davide} can assure ({\it a priori}) that in the atomistic region (compared to a full atomistic target data), only the first order of the probability distribution, i.e., the molecular number density, can be obtained with high accuracy. Higher orders of the distribution, atom-atom (or even molecule-molecule) radial distribution functions or three-body correlations, do not come automatically. Other relevant limitations are the violation of Newton's' third law and the fact that technically the method cannot avoid the integration of the atomistic phantom DOF's in the coarse-grained region. In fact such DOF's are part of the Hamiltonian, which means that in this method such DOF's are not phantom and that the coarse-grained zone is necessarily not defined by the coarse-grained
representation only. As it will be discussed later, the integration of the phantom DOF's in standard AdResS can be avoided and this would lead to a better performance of the code in terms of computational resources required. 
Nevertheless, the approach of \cite{davide}  represents the closest (somehow ``first order'') procedure for a truly adaptive Hamiltonian procedure with the major, advantage of the possibility of performing adaptive Monte Carlo simulations \cite{davide2}. 
A similar problem can be found in an approach recently proposed by Wagoner and Pande \cite{wagoner}. The method id based on the coupling of a full atomistic region to a continuum via an intermediate region where particles are inserted and removed via Monte Carlo moves biased by nonequilibrium paths. Interestingly, the high resolution region can change shape and adapt according to the evolution of the system. However, despite an analytic expression of the probability distribution of the system is reported, the only check done is about its validity at the first order, that is, at the level of the particle number density. This is not sufficient to prove the robustness of the method, it must be proven that higher orders can be reproduced. Moreover, the direct coupling to the continuum, differently from the adaptive methods above, based on  the treatment of large reservoir of coarse-grained particles, implicitly implies that particle number fluctuations are biased and thus the thermodynamics is artificial and can be properly described only if it is imposed externally.\\
The opinion of the authors is that a global Hamiltonian approach is not really required if it cannot be defined in a rigorous way; where rigorous means that it must satisfy all the basic requirements of an adaptive system. Alternative procedures based on fundamental principles of thermodynamics and statistical mechanics, if properly described and tested, and, as long as advantages and limitations are both reported, should be more welcome than artificial Hamiltonians. However, as stated before, it is our opinion that the cultural barrier in the community of molecular simulation is, at the current stage, too high to allow for such developments. In this perspective we have written this chapter; we describe the principles, gives our motivations and report the limitations and open problems of our method. 
\section{Recent Applications}
\subsection{How to run AdResS in your computer: Technical Description}
Before starting any AdResS simulation, the coarse-grained potentials have to be specified first. This is done in VOTCA 
 simulation package \cite{votca} by using Iterative Boltzmann inversion \cite{ibi} or Force matching technique. In the latest version of AdResS (GC-AdResS) the derivation of the coarse-grained potential can be avoided by specifying any spherical potential (for simple molecular liquids).
 To calculate the thermodynamic force and thus, the excess chemical potential defined in Eq. \ref{finalmu}, we use the 
 AdResS version of GROMACS \cite{gromacs} together with VOTCA. In the settings.xml file in VOTCA, the method is specified
 in the inverse section:
 \begin{verbatim}
   <method>tf</method> 
 \end{verbatim}
 For each interaction, some additional options have to be specified in the settings.xml file. 
 The options are given in units of Gromacs, which means nm, kJ/mol, K and ps. 
 To specify in which region the thermodynamic force should be nonzero, the 'min'
 and 'max' properties are used. A smoothing function proportional to $cos^{2}(r)$ is used to make the
 force go smoothly to zero at the region specified by 'min' and 'max'. Additionally a ’tf’ section is
 needed for each interaction type. Below is the section of setting.xml file from a urea/water simulation.
 \begin{verbatim}
  <non-bonded>
    <!-- name of the interaction -->
    <name>SOL</name>
    <!-- thermodynmic force: apply in this range -->
    <min>.45</min>
    <max>3.30</max>
    <step>0.01</step>
    <!--spline:offset beyond the hyrbid zone to take into account for fitting the spline-->
    <inverse>
      <!-- target distribution -->
      <target>dens.SOL.xvg</target>
      <!-- name of the table for gromacs -->
      <gromacs>
        <table>tabletf_CMW.xvg</table>
      </gromacs>
      <tf>
        <spline_start>.4</spline_start>
        <spline_end>3.35</spline_end>
        <spline_step>0.42142857142857142857</spline_step>
        <molname>SOL</molname>
        <prefactor>0.04</prefactor>
    </tf>
    </inverse>
  </non-bonded>

  <non-bonded>
    <!-- name of the interaction -->
    <name>UREA</name>
    <!-- thermodynmic force: apply in this range -->
    <min>.45</min>
    <max>3.30</max>
    <step>0.01</step>
    <!--spline:offset beyond the hyrbid zone to take into account for fitting the spline-->
    <inverse>
      <!-- target distribution -->
      <target>dens.UREA.xvg</target>
      <gromacs>
        <table>tabletf_CMC.xvg</table>
      </gromacs>
      <tf>
        <spline_start>.4</spline_start>
        <spline_end>3.35</spline_end>
        <spline_step>0.42142857142857142857</spline_step>
        <molname>UREA</molname>
        <prefactor>0.04</prefactor>
    </tf>
    </inverse>
  </non-bonded>
 \end{verbatim}
 Spline interpolation is used to smooth the force as the density highly fluctuates. To specify the 
 spline interpolation range the 'spline$\backslash_{-}$start' and 'spline$\backslash_{-}$end' parameters are used. These can define a larger region than that between
 min and max as it is sometimes useful to extend the spline fit for numerical stability. The parameter 'spline$\backslash_{-}$step' specifies the bin width of the fit grid. The 'molname' specifies the molecule used for 
 calculating the density. The prefactor $\frac{1}{\rho_{o}^{2}\kappa_{T}^{at}}$ appearing in Eq. \ref{kappa} is specified
 in the 'prefactor' field. It is seen that, if a very large prefactor is taken, then the thermodynamic force as well as the density 
 profile in the transition region strongly fluctuates, instead, with a small prefactor, the time taken 
 to obtain a converged thermodynamic force is large, thus few attempts must be done to find a reasonable compromise. In the urea/water example, a prefactor of 0.04 for urea, and a prefactor 
 of 0.08 for water was found to be a good choice. 
 A target density file has to be specified for each interaction type, in most cases this will contain a flat
 density profile at the equilibrium density $\rho_{o}$. In addition to these fields, there are some other parameters that are specified in the 
 inverse section. For each component, an initial guess of the thermodynamic force has to be provided in files 'tabletf$\backslash_{-}$CMC.xvg' (CMC is the name 
 of coarse-grained urea site) and 'tabletf$\backslash_{-}$CMW.xvg' (CMW is the name of coarse-grained water site). An efficient initial strategy is to add the 
 thermodynamic forces obtained from the simulation of pure components. In case, it is not available, then an initial guess with zero thermodynamic 
 force is the most suitable option. There are some other parameters that are written in the inverse section of the settings.xml file. 
 \begin{verbatim}
   <!-- general options for inverse script -->
  <inverse>
    <!-- 300*0.00831451, 300K in gromacs units -->
    <kBT>2.49435</kBT>
    <!-- use gromacs as simulation program -->
    <program>gromacs</program>
    <!-- gromacs specific options -->
    <gromacs>
      <!-- skip so many ps before calculation -->
      <equi_time>200</equi_time>
      <!-- grid for table*.xvg !-->
      <table_bins>0.01</table_bins>
      <!-- extend the tables to this value -->
      <table_end>5.83452</table_end>
    </gromacs>
    <!-- these files are copied for each new run -->
    <filelist>spc.adress.itp urea.adress.itp grompp.mdp topol.top table.xvg
      index.ndx table_CMW_CMW.xvg table_CMW_CMC.xvg table_CMC_CMC.xvg</filelist>
    <!-- do so many iterations -->
    <iterations_max>20</iterations_max>
    <!-- ibi: inverse biltzmann imc: inverse monte carlo tf: thermody force -->
    <method>tf</method>
  </inverse>
</cg>
 \end{verbatim}
 This section consists of parameters that are specific to gromacs mdrun. The files 'table$\backslash_{-}$CMW$\backslash_{-}$CMW.xvg', 'table$\backslash_{-}$CMW$\backslash_{-}$CMC.xvg' and 'table$\backslash_{-}$CMC$\backslash_{-}$CMC.xvg'
 consist of tabular coarse-grained potentials acting between water-water, water-urea and urea-urea coarse-grained molecules. The 'iterations$\backslash_{-}$max' command specifies
 the number of iterations required for thermodynamic force calculation. 
 The starting configuration has to be modified by adding a virtual coarse-grained site to all the molecules. For example, for the urea/water system, 
 the starting configuration would be the following:
 \begin{verbatim}
    1  SOL   OW    1   8.107   0.355   2.902 -0.0176  0.0989 -0.2993
    1  SOL  HW1    2   8.039   0.405   2.848 -0.2112  0.1880  0.0180
    1  SOL  HW2    3   8.131   0.270   2.855 -0.7787 -0.2104 -0.1498
    1  SOL  CMW    4   8.105   0.353   2.896 -0.0710  0.0866 -0.2732
     .
     .
     .
    2550 UREA   C110442   3.061   0.817   2.791  0.4370  0.0045  0.4733
    2550 UREA   O210443   3.019   0.936   2.780  0.0212 -0.1328  0.5610
    2550 UREA   N310444   3.001   0.706   2.741 -0.3024  0.4128  0.4390
    2550 UREA   H410445   3.039   0.614   2.736 -1.0670  0.1461 -2.0020
    2550 UREA   H510446   2.916   0.715   2.688  0.7484 -1.8617 -1.8763
    2550 UREA   N610447   3.182   0.798   2.847  0.7348 -0.0476 -0.1709
    2550 UREA   H710448   3.217   0.705   2.857  0.1912 -0.4974 -2.0263
    2550 UREA   H810449   3.244   0.874   2.865  1.5616 -0.0895 -2.5464
    2550 UREA  CMC10450   3.067   0.812   2.789  0.2179  0.0120  0.1647
 \end{verbatim}
  The virtual sites (CMC and CMW) are created at the center of mass of the molecule. This requires the 
  topology files (topol.top) to be modified by adding a 'virtual$\backslash_{-}$sites' field. 
  \begin{verbatim}
   [ atoms ]
;   nr    type   resnr  residu    atom    cgnr  charge  mass
     1     CUrea    1    UREA      C1       1   0.142  11.999
     2     OUrea    1    UREA      O2       1  -0.390  15.999
     3     NUrea    1    UREA      N3       1  -0.542  13.999
     4      H       1    UREA      H4       1   0.333   1.008
     5      H       1    UREA      H5       1   0.333   1.008
     6     NUrea    1    UREA      N6       1  -0.542  13.999
     7      H       1    UREA      H7       1   0.333   1.008
     8      H       1    UREA      H8       1   0.333   1.008
     9     CMC      1    UREA     CMC       2   0.000

   [ virtual_sitesn ]
   ; Site funct atom indexes
     9      2     1 2 3 4 5 6 7 8
     
   [ atoms ]
;   nr   type  resnr residue  atom   cgnr     charge       mass
     1     OW      1    SOL     OW      1      -0.82   15.99940
     2      H      1    SOL    HW1      1       0.41    1.00800
     3      H      1    SOL    HW2      1       0.41    1.00800
     4    CMW      1    SOL    CMW      2     0.0000

   [ virtual_sites3 ]
   ; Site from funct a d
    4 1 2 3 1 0.05595E+00 0.05595E+00
\end{verbatim}
  Also the 'atomtypes' and 'nonbon$\backslash_{-}$params' have to be specified separately for the coarse-grained sites. 
  \begin{verbatim}
   [ atomtypes ]
   ;name  mass        charge    ptype  C6             C12           ; sigma     epsilon
    CMW    0.00000     0.0000    V      0.000000E+00   1.000000E+00
    CMC    0.00000     0.0000    V      0.000000E+00   1.000000E+00
   [ nonbond_params ]
    ; i   j   func c6             c12
     CMW  CMW  1    0.000000E+00   1.000000E+00
     CMW  CMC  1    0.000000E+00   1.000000E+00
     CMC  CMC  1    0.000000E+00   1.000000E+00
  \end{verbatim}
In the GROMACS parameters file grompp.mdp, there are some additional parameters that are required with respect to a standard atomistic simulation. 
 For performing an AdResS simulation, 'adress$=$yes' has to be specified in the grompp.mdp file.
 \begin{verbatim}
  ; AdResS parameters
   adress = yes 
   adress_type              = xsplit
   adress_const_wf          = 1
   adress_ex_width          = 0.5
   adress_hy_width          = 2.75
   adress_ex_forcecap       = 0
   adress_interface_correction = thermoforce
   adress_site              = com
   adress_reference_coords  = 6.0 1.44344 1.44258
   adress_tf_grp_names      = CMW CMC
   adress_cg_grp_names      = CMW CMC
   adress_do_hybridpairs    = no
 \end{verbatim}
 There are two ways, in which one can construct the explicit and transition region in AdResS: a spherical region ('sphere' in 'adress$\backslash_{-}$type') 
 or a rectangular region ('xsplit' in 'adress$\backslash_{-}$type'). 'adress$\backslash_{-}$ex$\backslash_{-}$width' and 'adress$\backslash_{-}$ex$\backslash_{-}$width' specify the size of explicit region and 
 transition region respectively. If the thermodynamic force is used for the pressure correction, then it has to be specified in 'adress$\backslash_{-}$interface$\backslash_{-}$correction'
 and 'adress$\backslash_{-}$reference$\backslash_{-}$coords' specifies the coordinates of the center of explicit region. 
\subsection{Liquids and Mixture}
\begin{figure}
\center
\includegraphics[width=0.475\textwidth]{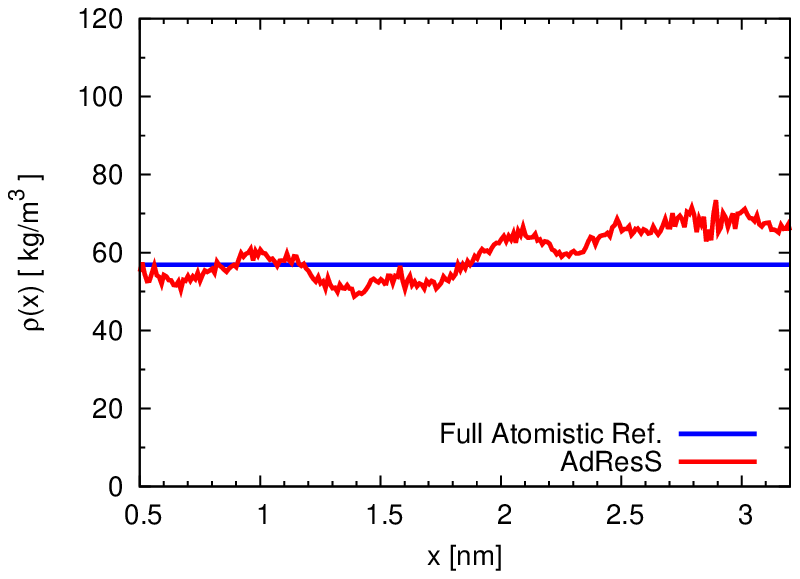}\\
\includegraphics[width=0.475\textwidth]{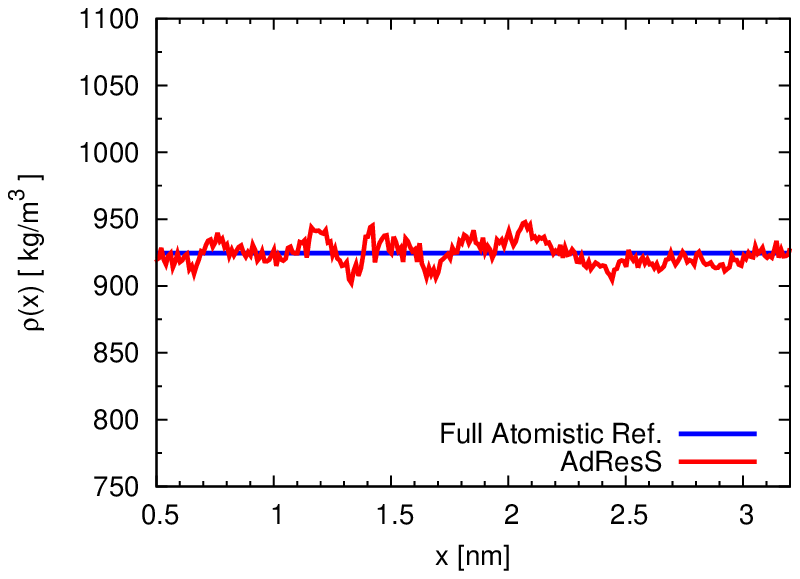}
\caption{Top: Molecular density profile in $\Delta$ for Urea/water mixture.\label{urea-water}}
\end{figure}

\begin{figure}
\center
\includegraphics[width=0.475\textwidth]{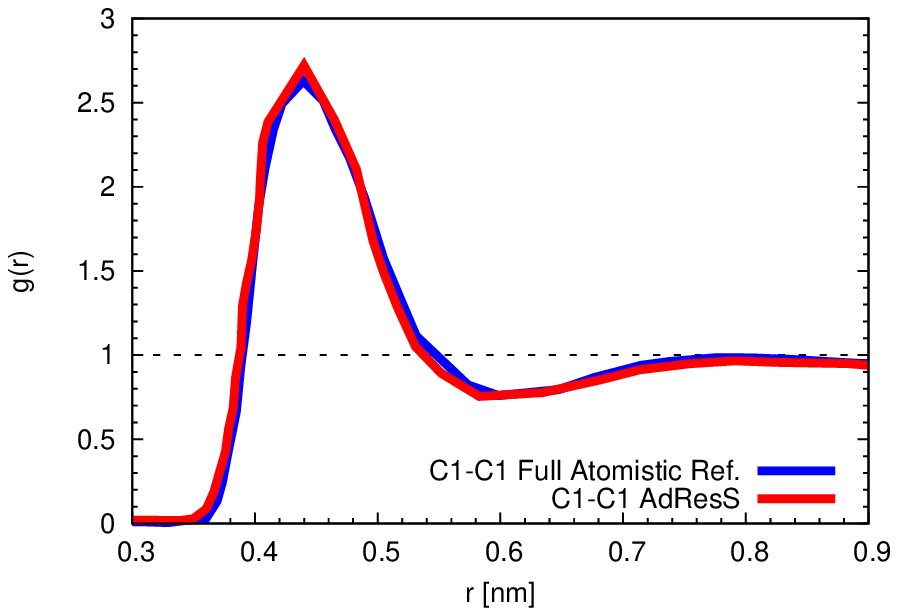}\\
\includegraphics[width=0.475\textwidth]{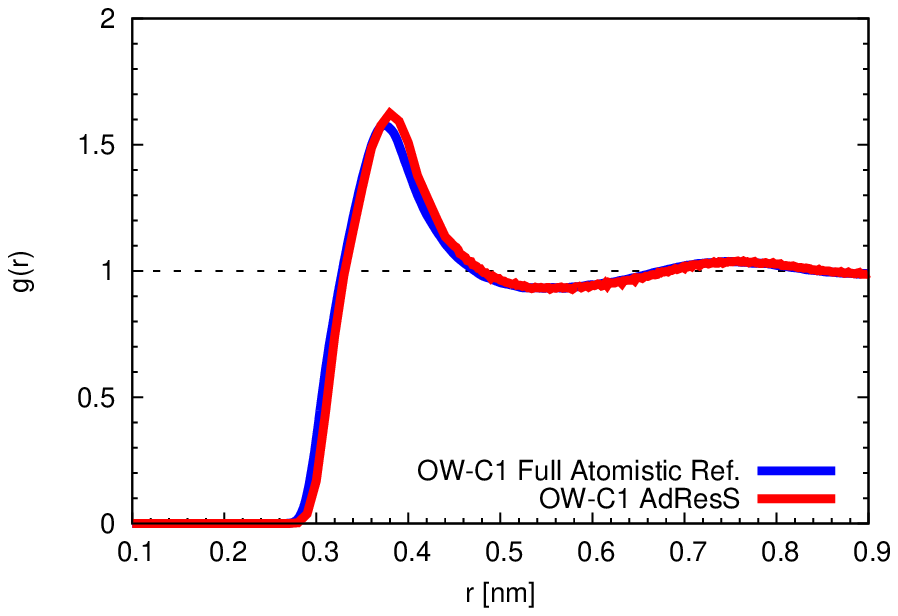}\\
\includegraphics[width=0.475\textwidth]{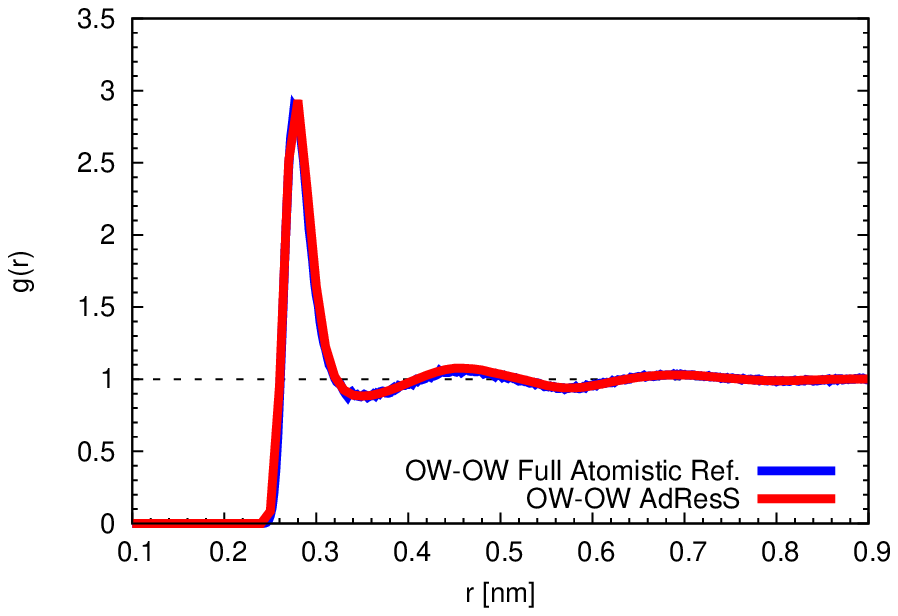}
\caption{Top: Urea-Urea radial distribution function (carbon-carbon); middle: the same plot for Urea-water (carbon-oxygen); bottom: for water-water (oxygen-oxygen). \label{gr}}
\end{figure}

In this section, as an example, we report a study of different liquids and mixtures. In particular we focus on the calculation of the excess chemical potential, $\mu^{ex}$. Such a quantity is of particular interest for understanding the capability of the system to act as a solvent, a more detailed discussion can be found in Ref.\cite{ourpaper}.  Table \ref{tablecp}
shows the values obtained from AdResS simulation, along with the values obtained by using thermodynamic integration 
approach and the values available in literature \cite{vang,nico}. 
\begin{table}[htpb]
\begin{center}
\begin{tabular}{ccccc}
%\begin{tabular}{cccc}
\hline \hline
 Liquid component & Mole fraction of solute & GC-AdResS & TI full atomistic & Experiment \\
\hline
methane  & -- & $-4.6 \pm 0.1$  & $-5.2 \pm 0.1$ & -- \\
ethane   & -- & $-8.2 \pm 0.3$  & $-8.8 \pm 0.1$ & -- \\
propane  & -- & $-8.5 \pm 0.1$ & $-9.5 \pm 0.2$ & -- \\
methanol  & -- & $-20.1 \pm 0.1$ & $-20.6 \pm 0.4$ & $-20.5$ \cite{vang}  \\
DMSO & -- & $-32.2 \pm 0.3$ & $-34.7 \pm 0.7$ & $-32.2$ \cite{dmso}  \\
methanol in methanol/water mixture & 0.01 & $-18.1 \pm 0.2$ & $-19.7 \pm 0.2$ & -- \\
methane in methane/water mixture & 0.006 & $9.1 \pm 0.1$  & $8.5 \pm 0.2$ & -- \\
urea in urea/water mixture & 0.02 & $-56.1 \pm 0.6$ & $-58.2 \pm 0.5$ & $-57.8 \pm 2.5$ \cite{urea} \\
ethane in ethane/water mixture & 0.006 & $7.2 \pm 0.2$ & $7.4 \pm 0.3$ & -- \\ 
TBA in water/TBA mixture & 0.001 & $-19.5 \pm 0.3$ & $-20.8 \pm 0.6$ & $-19.0$ \cite{nico} \\
DMSO in DMSO/water mixture & 0.01 & $-31.4 \pm 0.5$ & $-33.2 \pm 0.3$ & -- \\
TBA in TBA/DMSO mixture & 0.02 & $-24.8 \pm 0.4$ & $-24.0 \pm 0.5$ & -- \\
\hline \hline
\end{tabular}
\caption{The excess chemical potential of different liquids and mixtures calculated from GC-AdResS and thermodynamic integration (TI) of full atomistic simulations. The units for $\mu^{ex}$ are $kJ/mol$.}
\label{tablecp}
\end{center}
\end{table}
Results show that indeed the accuracy of AdResS for $\mu^{ex}$ is rather satisfactory; moreover it offers the advantage that, since the calculation of $\mu^{ex}$ corresponds to the equilibration of the system (see section\ref{musec}), every time an AdResS is performed the chemical potential is automatically obtained.
Fig.\ref{urea-water} shows the density profiles for a urea/water mixture in the transition region. The action of the thermodynamic force and 
thermostat in the transition region, that is the essential ingredients of $\mu^{ex}$, ensure that the density is sufficiently close to the reference full atomistic density. Due to the accuracy of the density in the transition region, the density in the atomistic region reproduces in a very accurate way the target one (not shown), in  addition, in Fig.\ref{gr} it is shown that the 
various radial distribution functions in the atomistic region in the urea/water system calculated from AdresS are in very good agreement with the target ones. The essential thermodynamic ($\mu^{ex}$) and structural ($\rho; g(r)$) quantities are properly reproduced.
%\label{table}
%\end{center}
%\end{table}
\section{Perspectives}
\subsection{Extension to (some) quantum problems: Path Integral AdResS}
\begin{figure}[h!]
\centering
\includegraphics[width=0.45\textwidth]{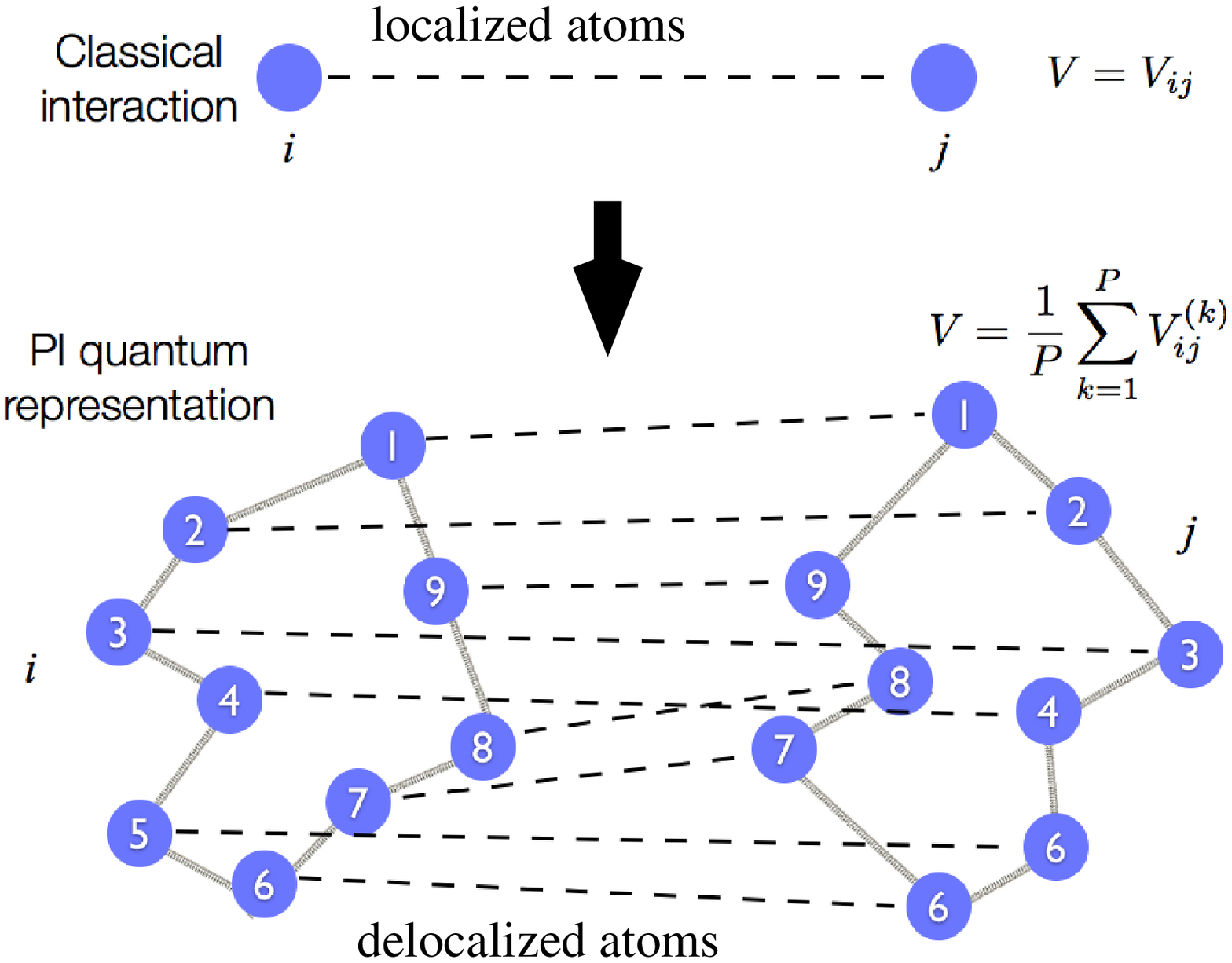}
\caption{\small{Schematic representation of the mapping from a classical description of the interaction between atom $i$ and atom $j$ (through the potential $V_{ij}$) to the path integral/polymer ring description. $P$ indicates the number of beads of the polymer, which in principle should tend to infinity. However for standard systems as been proved that 40 beads per atom is a satisfactory approximation.} 
\label{polrings}}
\end{figure}

The adaptive coupling of quantum and classical resolution is a rather delicate question, in fact the problem is that we are interfacing not only systems with a different number of degrees of freedom, but above all the equations governing their evolution are different: deterministic on the classical side (Newton's equation), probabilistic on the other (Schr\"{o}dinger equation), and thus the idea of adaptivity necessarily implies some sort of conceptual discontinuity. For electrons this is a rather difficult problem in fact it is not clear how to switch on and off an electron, which by definition is a non localized particle. One must also keep in mind that the electronic properties in the quantum region of the adaptive box, must be the same as if the whole system was treated at quantum level.
Practical cases where the adaptive idea is employed in the context of electronic structure calculations do not fulfill the requirement above, since the main properties of interest are not the electronic ones. In fact in this case the electronic structure is used only as an on-the-fly good approximation to refine the classical force field \cite{ensing2,devita,paesani}. A rigorous solution has been found for a class of quantum systems, namely those where light atoms are present. In fact light atoms, e.g. hydrogens (protons), even at room temperature display non negligible effects of spatial delocalization; such a delocalization can be described through the formalism of the path integral of Feynman \cite{hibbs}. While in the classical description two atoms are represented by two spheres interacting with a given potential, in the path integral formalism two atoms are mapped onto ''effective'' classical polymer rings and the interaction distributed among pairs of beads with the same index (bead one or polymer one with bead one of polymer two etc etc) (see Fig.\ref{polrings}). 
The fluctuation of the shape of the polymer ring in space describes the quantum delocalization of the atom and the statistical properties of such systems can be properly described via classical molecular dynamics (see e.g. \cite{tuckermann}). For the adaptive approach this is an ideal situation; in fact interfacing a quantum and a classical description is equivalent to interface two regions with a different number of ''effective'' classical degrees of freedom, and thus the computational apparatus developed so far can be applied straightforwardly.  

\begin{figure}[h!]
\centering
\includegraphics[width=0.5\textwidth]{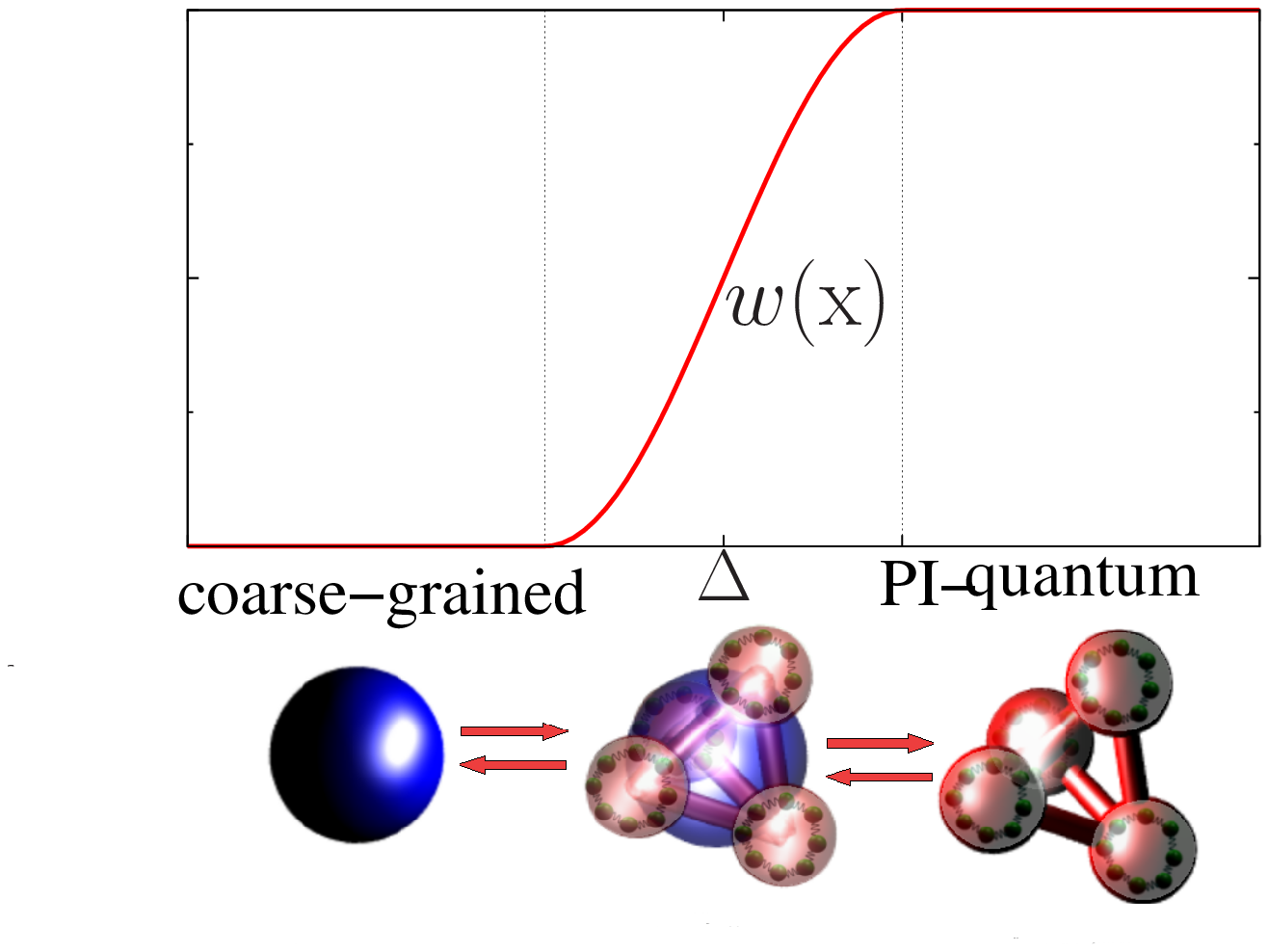}
\caption{\small{Schematic representation of the adaptive resolution for the tetrahedral molecule which in our work is usually employed to test the adaptive principles. In the quantum region, (right) each atom is represented by a polymer ring of 30 beads, in the $\Delta$ region the molecule has hybrid coarse-grained/quantum resolution and in the coarse-grained region the molecule is represented by an effective spherical model obtained from a full path integral simulation.} 
\label{pathmol}}
\end{figure}

\begin{figure}[h!]
\centering
\includegraphics[width=0.95\textwidth]{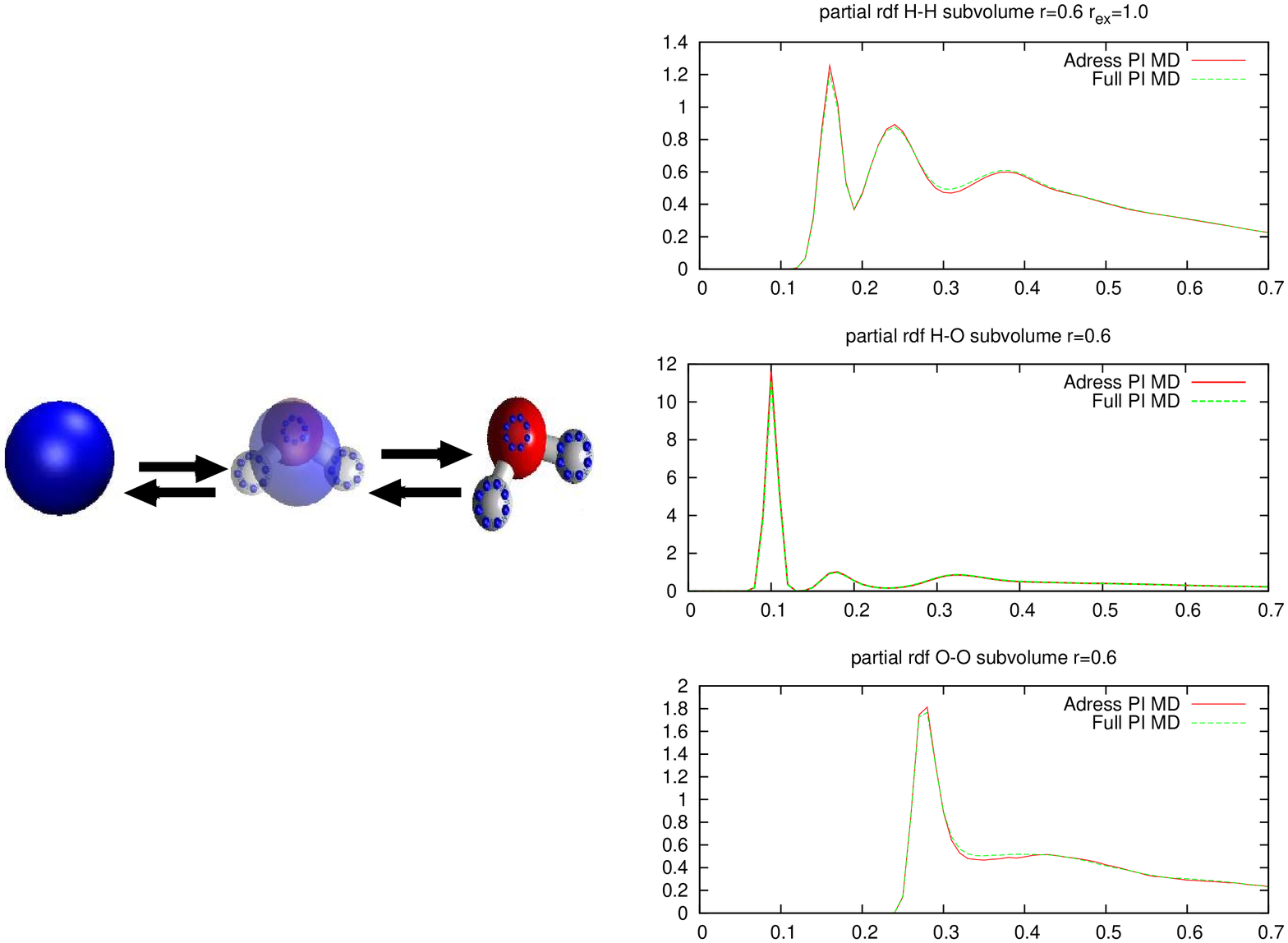}
\caption{\small{Left: Schematic representation of the path integral adaptive resolution for a water molecule. In the quantum region each atom is represented by a polymer ring of 40 beads, in the transition region the molecule acquire hybrid coarse-grained/quantum resolution and in the coarse-grained region the molecule is represented by an effective spherical model obtained from the full path integral simulation. Right: radial distribution functions for hydrogen-hydrogen (top), oxygen-hydrogen (middle), oxygen-oxygen (bottom). The $g(r)$'s are calculated in the quantum region of AdResS and are compared with the corresponding function calculated in an equivalently large subregion of a full path integral simulation. Results show that AdResS satisfactorily reproduces the reference data.} 
\label{wat-fif-pi}}
\end{figure}

\begin{figure}[h!]
\centering
\includegraphics[width=0.65\textwidth]{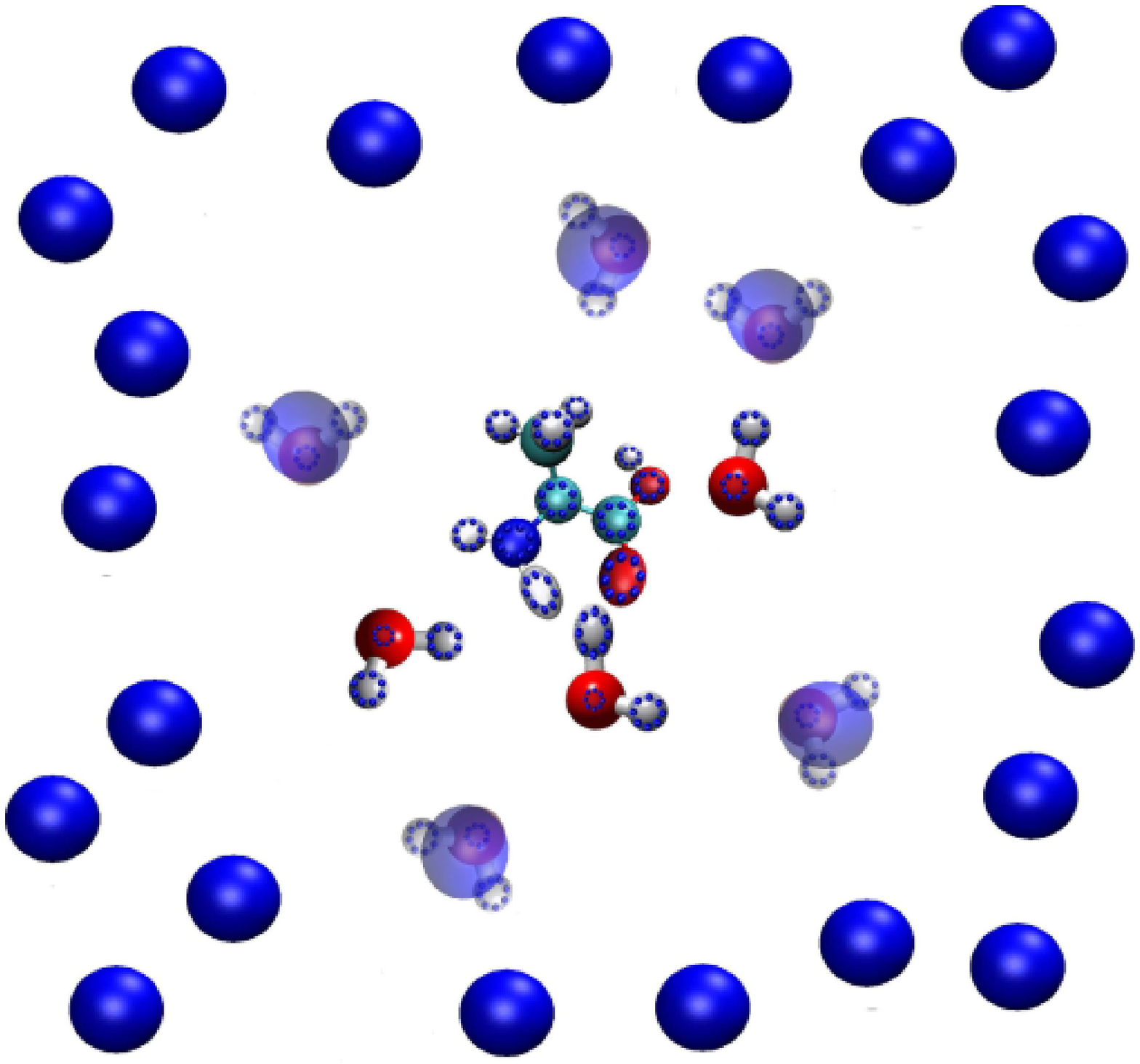}
\caption{\small{Pictorial representation of the solvation of a small biomolecule in water within the AdResS path integral scheme. Around the solute in the first solvation shell, hydrogen and oxygens are treated at path integral level, thus the effects of spatial delocalization of the hydrogen is properly described; far way from the solute, water is represented by classical spherical particles. The transition from one representation to another occurs as in standard classical AdResS via a space dependent hybrid molecular representation.} 
\label{PI-bio-mol}}
\end{figure}

Indeed this was shown to be the case for the liquid of tetrahedral molecules, where in the quantum region each atom of the molecule is represented by a polymer ring with 30 beads while in the coarse-grained region the whole molecule is represented as a spherical coarse-grained object (see Fig.\ref{pathmol} and Ref.\cite{prlado}). 
Moreover a further test was made for a liquid of parahydrogen at extreme thermodynamic conditions where quantum effects play a major role and data from previous, full path integral, simulations were available as a reference. Also in this case the adaptive approach was able to reproduce the basic quantum properties of reference (see \cite{pccpado,jcpraff}).  More recently we have tested the approach for liquid water at room conditions using a water model specifically designed for path integral simulations \cite{paesanimod}. Fig.\ref{wat-fif-pi} shows the various radial distribution functions calculated in the high resolution region of AdResS and compared with the equivalent quantities  from a full path integral simulation: although one must consider such results at preliminary level, however they are definitively encouraging. Fritsch {\it et al.} have in the meanwhile developed a water model for path integral simulations which properly distinguishes between electronic and nuclear delocalization effects so that double counting of effects in the parameterization are avoided when the path integral approach is used; such model will be ideal for AdResS simulations \cite{davide-abinit} (for an extended discussion about path integral approach and AdResS see also \cite{sebastian-thesis}).
What is the utility of such an approach for simulating biological molecules in solution?\\
Fig.\ref{PI-bio-mol} pictorially shows the solvation of a biomolecule in water using AdResS with path integral. The flexibility of the hydrogen bond network may strongly depend on the spatial delocalization effects of the hydrogens which in turn may imply that the conformational properties of the molecule in solution can be very different from those obtained by standard atomistic simulation. The advantage of AdResS is that since it requires only a relatively small region at quantum level, the computational costs are not high and such simulations can be performed also by research groups which have got only standard (in house) computational resources; currently, full path integral simulation for such systems are possible only in presence of massive computational resources.
\subsection{On the dynamical properties of AdResS}
\begin{figure}[h!]
\centering
\includegraphics[width=0.6\textwidth]{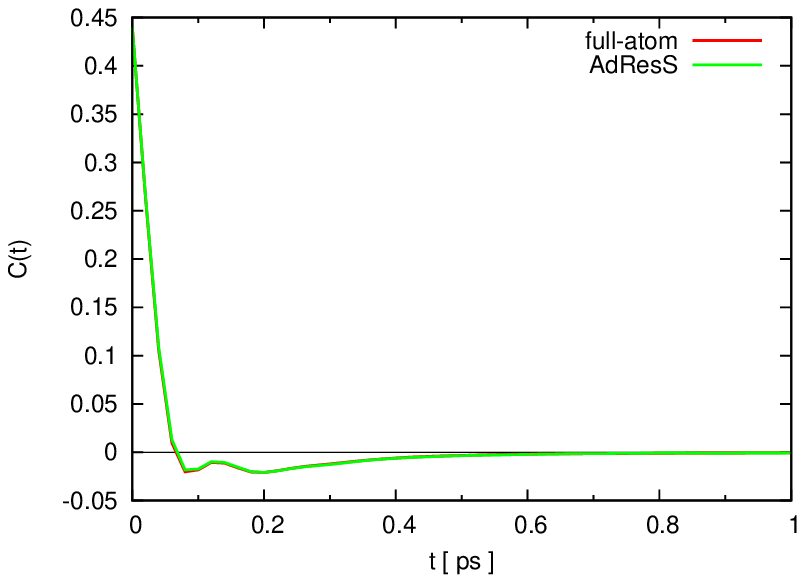}
\caption{\small{The  velocity auto-correlation function, $C(t) = \langle v(t)
\cdot v(0)\rangle$ plotted against time. In green the data obtained with AdResS, in red the data obtained from a full atomistic simulation.} 
\label{fig:velo-corr}}
\end{figure}

An interesting question regarding AdResS is the accuracy of the method in determining dynamical properties in the atomistic region. In fact, due to the perturbation to the dynamics introduced by the transition region, trajectories within the atomistic region can be treated, strictly speaking, only as tools for statistical sampling of configurations, rather than for their intrinsic dynamical properties. However, one may expect that, at least for short time scales, dynamical processes should be correctly described. In order to test the validity of dynamics in the atomistic region, as a preliminary test, we have calculated the velocity auto-correlation function, $C(t) = \langle v(t)
\cdot v(0)\rangle$ in an AdResS system (for more details, see Ref.~\cite{jinglong}), and compared it with the
the same quantity calculated in a full-atomistic system of reference.  
Technically, the full-atomistic system consists of 1728 SPC
water molecules in a cubic box of length 3.724 nm. The time step is
2.0 fs and the neighbor list, which is built with a cut-off of 0.9
nm, is updated every 10 steps. We use reaction field with Coulomb
cut-off 0.9 nm with dielectric constant of 80 in order to treat the
electrostatics in the atomistic region. The simulation lasts for 2.0 ns, and the first 0.5 ns are removed. The AdResS simulation box is
$7.529\times3.765\times3.765\ \textrm{nm}^3$ containing 3456 water
molecules. The system is split into atomistic, hybrid and
coarse-grained resolution in x-direction. The size of the atomistic
region is 1.0 nm and that of the hybrid region is 2.7 nm. The
thermodynamics force is applied to the system to ensure a uniform
density distribution in the system.  All other running parameters in
the AdResS simulation are the same as those in the full-atomistic
simulation of reference.  In the AdResS system, the velocity auto-correlation is
calculated only when molecules are in the atomistic region.
Results are reported in Fig.~\ref{fig:velo-corr} where is shown a highly satisfactory consistency between the
AdResS simulation and full-atomistic reference simulation.
In this preliminary test we have considered only one system with a given size of the atomistic region, in future work we will perform a systematic study of $C(t)$ as a function of the size of the atomistic region~\cite{jinglong}; this will allow us to conclude about the validity of a basic dynamical property as a function of the atomistic size and thus to suggest the limitation of AdResS in describing dynamics.
\subsection{Further Computational Optimization}
AdResS has been implemented, so far, in different molecular dynamics codes, namely ESPResSo~\cite{jungsim}, ESPResSo++~\cite{espressopp} and Gromacs~\cite{gromacs}. 
All these implementations follow the ``double representation trick''~\cite{jungsim}, which means the atomistic identities of the molecules are kept even in the coarse-grained zone.
This approach leads to few computational advantages regarding the equilibration of degrees of freedom once the molecules enters into the hybrid zone.
On the other hand, the force calculation, which is the computational most expensive part, is performed only on the physical relevant identities (degrees of freedom).
The simplification of force calculation allows AdResS to provide a performance which, in terms of computational costs, is close to that of a pure coarse-grained simulation (for a small atomistic region). 
Moreover the atomistic as well as the coarse-grained zone of the simulation can be view as a standard MD simulation for which common performance tunings can be applied.
For instance the Gromacs implementation~\cite{gromacs} benefits from this fact and uses the highly optimized and vectorized force kernel in these two zones.
Using the Gromacs version 4.5.1 a speedup of a factor 4 compared to full atomistic simulations has been reported for system of mixtures~\cite{debashish1,debashish2}.
However in the successive version of GROMACS 4.6.1 the performance of atomistic simulations was highly improved while the corresponding implementation of AdResS is still not optimized. As a consequence, at the current state, AdResS can only assure a speed up between 2.0 and 2.5 for large systems (30000 molecules) compared to full atomistic simulations.  
In perspective, there is large room for improvements regarding the technical implementation in multiple directions.
First, the force calculations of the AdResS part could be vectorized, too.
Second, as the coarse-grained part is computationally less demanding than the rest of the simulation, the box needs to be divided into non-uniform domains among the different processors to decrease the load imbalance and the ``double representation trick'' can be avoided by removing the atomistic identity.
Third, the time step in the coarse-grained region can be increased due to the increased softness of the coarse-grained interactions.

\subsection{Possible extension to non equilibrium}
A possible extension of AdResS is that of treating systems which are not in equilibrium. A possible path for such an extension is to merge AdResS with a dynamical approach to the nonequilibrium molecular dynamics (D-NEMD) as developed in Refs.~\cite{ciccotti1975direct,
  ciccotti1979thought}. In brief, such method considers the the response of systems in equilibrium to non-equilibrium perturbations following Onsager's regression hypothesis \cite{Onsager1931a,Onsager1931b}. The hypothesis states that, up to a certain extent, external perturbations can be treated as (large) fluctuations within the  equilibrium statistics of the system, and that the system's non-equilibrium response can be compared with fluctuations slowly returning back to equilibrium. An observable, $O(t)$, in a system subject to a perturbation which drives it out of equilibrium, can be calculated as:
\begin{align}\label{eqn:nemd-algorithm}
  O(t) = \int d\vect x\, \hat O(\vect x(t)) \rho(\vect x, 0), 
\end{align}
where $\vect x$ is the phase space variable of the system. $\hat
O(\vect x)$ is the microscopic observable, which is a only a function
of the phase space position.  $\rho(\vect x, 0)$ is the initial
(equilibrium) probability density.
The advantage is that one uses the equations and thermodynamic relations of equilibrium, that is, one uses equilibrium simulations to derive an ensemble of initial configuration along each of which a non equilibrium simulation is performed. $O(t)$ is then calculated averaging over the non equilibrium trajectories.  
The method has been developed for Canonical ensembles and strongly relies on the Liouville equation. In order to extend such an approach to (the atomistic region of) AdResS there are two major points that must be treated: (a) the method must be extended to the Grand Canonical case, which implicitly requires the extension of Liouville equation to open systems, thus with varying number of particles;
(b) the thermostat used in AdResS for the removing/insertion of degrees of freedom should be such that it does not create artifacts by adsorbing the effects of the external perturbation. A proper solution of problem (a) requires a major conceptual step forward, instead point (b) has already been addressed in some recent work, by defining of a thermostat that acts only outside the region where the perturbation is applied \cite{non-eq-han}. 
Once problem (a) will be solved, then the application of AdResS to non equilibrium systems will be possible (for more details, see Ref.~\cite{ejp}).
\section*{ Conclusion} 
We have reported the basic principles and the technical aspects of AdResS. Its conceptual evolution, from intuitive physical principles to a more rigorous formalization in terms of Grand Canonical-like set up, was extensively discussed. A list of open problems was given, in particular the fact that the conditions in the transition region, are only necessary but not sufficient. Moreover, the current computational gain, is not more than a factor two, compared with equivalent full atomistic simulations. Although such a gain is not trivial, expecially for large biological systems, however it can be highly improved upon technical optimization of the code; this aspect is probably the most relevant for applications.

\section*{Acknowledgments}
This work was supported by the Deutsche Forschungsgemeinschaft (DFG) with the Heisenberg grant provided to L.D.S (grant code DE 1140/5-1) and with its associate DFG grants for A.G.(grant code DE 1140/7-1) and for H.W. (grant code DE 1140/4-2). CJ was funded by a LANL Director’s fellowship.


\begin{thebibliography}{24}
\expandafter\ifx\csname natexlab\endcsname\relax\def\natexlab#1{#1}\fi
\expandafter\ifx\csname url\endcsname\relax
  \def\url#1{\texttt{#1}}\fi
\expandafter\ifx\csname urlprefix\endcsname\relax\def\urlprefix{URL }\fi

\bibitem{jcp}
M.Praprotnik, L.Delle Site, and K.Kremer
\newblock Adaptive resolution molecular dynamics simulation: Changing the degrees of freedom on the fly.
\newblock \emph{J.Chem.Phys.} \textbf{123}, 224106 (2005).

\bibitem{prl1}
S.Fritsch, S.Poblete, C.Junghans, G.Ciccotti, L.Delle Site and K.Kremer
\newblock Adaptive Resolution Molecular Dynamics Simulation Through Coupling to an Internal Particle Reservoir.
\newblock \emph{Phys.Rev.Lett.}  \textbf{108} 170602 (2012).

\bibitem{prx}
H.Wang, C.Hartmann, C.Sch\"{u}tte and L.Delle Site
\newblock Grand-canonical-like molecular-dynamics simulations by using an adaptive-resolution technique.
\newblock \emph{Phys.Rev.X}  \textbf{3} 011018 (2013).

\bibitem{entropy}
L.Delle Site
\newblock What is a Multiscale Problem in Molecular Dynamics?
\newblock \emph{Entropy}  \textbf{16} 23-40 (2014).

\bibitem{Laio:2002}
M.C.Colombo, L.Guidoni, A.Laio, A.Magistrato, P.Maurer, S.Piana, U.Rohrig, K.Spiegel, M.Sulpizi, J.VandeVondele, M.Zumstein, U.R\"{o}thlisberger 
\newblock Hybrid QM/MM Car-Parrinello simulations of catalytic and enzymatic reactions.
\newblock \emph{CHIMIA}  \textbf{56} 13 (2002).

\bibitem{jungsim}
C.Junghans and S.Poblete
\newblock A reference implementation of the adaptive resolution scheme in ESPResSo.
\newblock {\em Comp.Phys.Comm.}  \textbf{181} 1449 (2010).

\bibitem{espressopp}
J.D.Halverson,T.Brandes, O.Lenz, A.Arnold, S.Bevc, V.Starchenko, K.Kremer, T.Stuehn, D.Reith
\newblock ESPResSo++: A modern multiscale simulation package for soft matter systems
\newblock \emph{Comp.Phys.Comm.} \textbf{184} 1129 (2012).

\bibitem{jctc}
H.Wang, C.Sch\"{u}tte and L.Delle Site
\newblock Adaptive Resolution Simulation (AdResS): A smooth thermodynamic and structural transition from atomistic to coarse grained resolution and vice versa in a Grand Canonical fashion.
\newblock \emph{J.Chem.Th.Comp.}  \textbf{8} 2878 (2012).

\bibitem{prefirst}
M.Praprotnik, L.Delle Site and K.Kremer
\newblock Adaptive Resolution Scheme (AdResS) for Efficient Hybrid Atomistic/Mesoscale Molecular Dynamics Simulations of Dense Liquids.
\newblock \emph{Phys.Rev.E} \textbf{73} 066701 (2006).

\bibitem{jcppol}
M.Praprotnik, L.Delle Site and K.Kremer
\newblock A macromolecule in a solvent: Adaptive resolution molecular dynamics simulation.
\newblock \emph{J.Chem.Phys.}  \textbf{126} 134902 (2007).

\bibitem{wat1}
M.Praprotnik, S.Matysiak, L.Delle Site, K.Kremer and C.Clementi
\newblock Adaptive resolution simulation of liquid water 
\newblock \emph{J.Phys.Cond.Matt.} \textbf{19} 292201 (2007).

\bibitem{wat2}
S.Matysiak, C.Clementi, M.Praprotnik, K.Kremer and L.Delle Site
\newblock Modeling Diffusive Dynamics in Adaptive Resolution Simulation of Liquid Water.
\newblock  \emph{J.Chem.Phys.}  \textbf{128} 024503 (2008).

\bibitem{jcpcover}
B.P. Lambeth, C.Junghans, K.Kremer, C.Clementi, and L.Delle Site
\newblock On the Locality of Hydrogen Bond Networks at Hydrophobic Interfaces.
\newblock \emph{J.Chem.Phys.}  \textbf{133} 221101 (2010).

\bibitem{ibi}
D.Reith, M.P\"{u}tz and F.M\"{u}ller-Plathe
\newblock Deriving effective mesoscale potentials from atomistic simulations.
\newblock \emph{J.Comp.Chem.} \textbf {24} 1624-1636 (2003).

\bibitem{votca}
V.Ru\"{u}hle, C.Junghans, A.Lukyanov, K.Kremer and D.Andrienko
\newblock Versatile object-oriented toolkit for coarse-graining applications.
\newblock \emph{J.Chem.Th.Comp.} \textbf{5} 3211-3223 (2009).

\bibitem{jpa}
M.Praprotnik, K.Kremer and L.Delle Site
\newblock Fractional dimensions of phase space variables: A tool for varying the degrees of freedom of a system in a multiscale treatment.
\newblock \emph{J.Phys.A:Math.Th.} \textbf{40} F281-F288 (2007).

\bibitem{preluigi}
L.Delle Site
\newblock Some fundamental problems for an energy-conserving adaptive-resolution molecular dynamics scheme. 
\newblock \emph{Phys.Rev.E} \textbf{76} 047701 (2007).


\bibitem{prefrac}
M.Praprotnik, K.Kremer and L.Delle Site
\newblock Adaptive molecular resolution via a continuous change of the phase space dimensionality.
\newblock  \emph{Phys.Rev.E} \textbf{75} 017701 (2007).

\bibitem{tint}
I. G. Tironi and W. F. van Gunsteren
\newblock A molecular dynamics simulation study of chloroform
\newblock \emph{Mol.Phys.} \textbf{83}   381-403 (1994).

\bibitem{ipm}
B.Widom
\newblock Some Topics in the Theory of Fluids
\newblock \emph{J.Chem.Phys.} \textbf{39} 2808 (1963).


\bibitem{jcpmu}
S.Poblete, M.Praprotnik, K.Kremer and L.Delle Site
\newblock Coupling different levels of resolution in molecular simulations.
\newblock \emph{J.Chem.Phys.} \textbf{132} 114101 (2010).


\bibitem{hanpaper}
H.Wang, C.Junghans and K.Kremer
\newblock Comparative atomistic and coarse-grained study of water: What do we lose by coarse-graining? 
\newblock \emph{Eur.Phys.Jour.E} \textbf{28} 221-229 (2009).

\bibitem{debashish1}
D.Mukherji, N.F.A.van der Vegt, K.Kremer and L.Delle Site
\newblock Kirkwood-Buff analysis of liquid mixtures in an open boundary simulation.
\newblock  \emph{J.Chem.Th.Comp.} \textbf{8} 375 (2012).

\bibitem{debashish2}
D.Mukherji, N.F.A.van der Vegt, K.Kremer 
\newblock Preferential solvation of triglycine in aqueous urea: An open boundary simulation approach.
\newblock \emph{J.Chem.Th.Comp.} \textbf{8} 3536 (2012).

\bibitem{ensing1}
B.Ensing, S.O. Nielsen, P.B. Moore, M.L. Klein, and M.Parrinello
\newblock Energy conservation in adaptive hybrid atomistic/coarse-grain molecular dynamics. 
\newblock  \emph{J.Chem.Th.Comp.} \textbf{3} 1100-1105 (2007).

\bibitem{ensing2}
R.E. Bulo, B.Ensing, J.Sikkema, and L.Visscher
\newblock Toward a Practical Method for Adaptive QM/MM Simulations.
\newblock \emph{J.Chem.Th.Comp.} \textbf{5} 2212 - 2221 (2009).


\bibitem{ensing3}
S.O. Nielsen, P.B. Moore, and B.Ensing
\newblock Adaptive multiscale molecular dynamics of macromolecular fluids. 
\newblock \emph{Phys.Rev.Lett.} \textbf{105} 237802 (2010).

\bibitem{prlcomm}
M.Praprotnik, S.Poblete, L.Delle Site and K.Kremer
\newblock Comment on: Adaptive multiscale molecular dynamics of macromolecular fluids
\newblock \emph{Phys.Rev.Lett.} \textbf{107} 099801 (2011). 

\bibitem{ensingagain}
B.Ensing, A.Laio, and S.O. Nielsen
\newblock Hamiltonian adaptive hybrid atomistic/coarse-grain molecular dynamics.
\newblock \emph{Publication series of the John von Neumann Institute for Computing} \textbf{46} (2013).

\bibitem{arxiv}
M.H.Peters
\newblock An extended liouville equation for variable particle number systems.
\newblock \emph{arXiv:physics} \textbf{9809039} (1998).

\bibitem{davide}
R.Potestio, S.Fritsch, P.Espanol, R.Delgado-Buscalioni, K.Kremer, R. Everaers and D.Donadio
\newblock Hamiltonian adaptive resolution simulation for molecular liquids. 
\newblock \emph{Phys.Rev.Lett.} \textbf{110}, 108301 (2013).

\bibitem{davide2}
R.Potestio, P.Espanol, R.Delgado-Buscalioni, R. Everaers, K.Kremer and D.Donadio
\newblock Monte Carlo adaptive resolution simulation of multicomponent molecular liquids.
\newblock  \emph{Phys.Rev.Lett.} \textbf{111}, 060601 (2013).

\bibitem{wagoner}
J.A.Wagoner and V.S.Pande
\newblock Finite domain simulations with adaptive boundaries: Accurate potentials and nonequilibrium movesets.
\newblock \emph{J.Chem.Phys.} \textbf{139} 234114 (2013).

\bibitem{gromacs}
S. Fritsch, C. Junghans and K. Kremer,
\newblock Structure formation of toluene around C60: Implementation of the Adaptive Resolution Scheme (AdResS) into GROMACS.
\newblock \emph{J.Chem.Th.Comp.} \textbf{8}, 398 (2012).

\bibitem{ourpaper}
A.Agarwal, H.Wang, C.Sch\"{u}tte and L.Delle Site
\newblock Chemical potential of liquids and mixtures via Adaptive Resolution Simulation
\newblock \emph{J. Chem. Phys.}\textbf{141} 034102 (2014).

\bibitem{vang}
D.P.Geerke and W.F. van Gunsteren
\newblock Force Field Evaluation for Biomolecular Simulation: Free Enthalpies of Solvation of Polar and Apolar Compounds in Various Solvents.
\newblock \emph{ChemPhysChem}  \textbf{7} 671 (2006).

\bibitem{nico}
M.E.Lee and N.F.A. van der Vegt
\newblock A new force field for atomistic simulations of aqueous tertiary butanol solutions.
\newblock \emph{J.Chem.Phys.} \textbf{122} 114509 (2005).

\bibitem{dmso}
J.R. Pliego Jr and J.M. Riveros
\newblock Gibbs energy of solvation of organic ions in aqueous and dimethyl sulfoxide solutions
\newblock \emph{Phys. Chem. Chem. Phys.} \textbf{4} 1622 (2002).

\bibitem{urea}
L.J. Smith and H.J.C. Berendsen and W.F. van Gunsteren
\newblock Computer Simulation of Urea?Water Mixtures:? A Test of Force Field Parameters for Use in Biomolecular Simulation
\newblock \emph{J. Phys. Chem. B} \textbf{108} 1065 (2004).

\bibitem{devita}
G.Csanyi, T.Albaret, M.C.Payne and A.De Vita 
\newblock ``Learn on the fly'': a hybrid classical and quantum-mechanical molecular dynamics simulation.
\newblock \emph{Phys.Rev.Lett.} \textbf{93}, 175503 (2004).


\bibitem{paesani}
K. Park, A. G\"{o}tz, R.C. Walker and F. Paesani
\newblock Application of adaptive QM/MM methods to molecular dynamics simulations of aqueous systems.
\newblock \emph{J.Chem.Th.Comp.} \textbf{8} 2868 (2012).

\bibitem{hibbs}
R.P.Feynman and A.R.Hibbs
\newblock Quantum Mechanics and Path Integrals.
\newblock McGraw-Hill Inc.

\bibitem{tuckermann}
\newblock Path Integration via Molecular Dynamics.
\newblock \emph{Publication series of the John von Neumann Institute for Computing} \textbf{10} (2002).

\bibitem{prlado}
A.B.Poma and L.Delle Site
\newblock Classical to Path-Integral Adaptive Resolution in Molecular Simulation: Towards a Smooth Quantum-Classical Coupling.
\newblock \emph{Phys.Rev.Lett.} \textbf{104} 250201 (2010).


\bibitem{pccpado}
A.B.Poma and L.Delle Site
\newblock Adaptive Resolution Simulation of Liquid Para-Hydrogen: Testing the robustness of the Quantum-Classical Adaptive Coupling.
\newblock \emph{Phys.Chem.Chem.Phys.} \textbf{13} 10510-10519 (2011).

\bibitem{jcpraff}
R.Potestio and L.Delle Site
\newblock Quantum Locality and Equilibrium Properties in Low-temperature Parahydrogen: A Multiscale Simulation Study.
\newblock \emph{J.Chem.Phys.} \textbf{136} 054101 (2012).

\bibitem{paesanimod}
F. Paesani, W. Zhang, T.E. Cheatham, III, D.A. Case and G.A. Voth
\newblock An accurate and simple quantum model for liquid water
\newblock  \emph{J.Chem.Phys.} \textbf{125} 18457 (2006).

\bibitem{davide-abinit}
S.Fritsch, R.Potestio, D.Donadio and K.Kremer
\newblock Nuclear Quantum Effects in Water: A Multi-Scale Study.
\newblock \emph{J.Chem.Th.Comp.} \textbf{10} 816 (2014).

\bibitem{sebastian-thesis}
S.Fritsch
\newblock Scale Bridging Concepts in Molecular Simulation: Coarse-Graining and Thermodynamic Coupling.
\newblock PhD Thesis in Physics, University of Mainz (2013).

\bibitem{jinglong}
J.Zhu, H.Wang, A.Agarwal and L.Delle Site
\newblock Equilibrium time correlation functions in open systems
\newblock \emph{arXiv:1412.0866} (2014).

\bibitem{ejp}
H. Wang and A.Agarwal
\newblock Adaptive Resolution Simulation in Equilibrium and Beyond
\newblock \emph{arXiv:1412.2340} (2014).

\bibitem{ciccotti1975direct}
G.Ciccotti G.and Jacucci
\newblock Direct computation of dynamical response by molecular dynamics: The mobility of a charged Lennard-Jones particle.
\newblock \emph{Phys.Rev.Lett.} \textbf{35} 789 (1975).

\bibitem{ciccotti1979thought}
G.Ciccotti, G. Jacucci and I.R.McDonald
\newblock ``Thought-experiments'' by molecular dynamics.
\newblock \emph{J.Stat.Phys.} \textbf{21} 1-22 (1979).

\bibitem{Onsager1931a}
L.~Onsager.
\newblock Reciprocal relations in irreversible processes. i.
\newblock \emph{Phys. Rev.} \textbf{37} 405 (1931).

\bibitem{Onsager1931b}
L.~Onsager.
\newblock Reciprocal relations in irreversible processes. ii.
\newblock \emph{\em Phys. Rev.} \textbf{38} 2265 (1931).

\bibitem{non-eq-han}
H.Wang, C.Sch\"{u}tte, G.Ciccotti and L.Delle Site
\newblock Exploring the conformational dynamics of alanine dipeptide in
solution subjected to an external electric field: A nonequilibrium
molecular dynamics simulation.
\newblock  \emph{J.Chem.Th.Comp.} \textbf{10} 1376 (2014).



\end{thebibliography}
\end{document}